\documentclass[aps, prb, twocolumn,superscriptaddress,floatfix,longbibliography]{revtex4-2}
\usepackage[utf8]{inputenc}
\usepackage{graphicx,amsfonts,amssymb,amsmath,mathrsfs,hyperref,bm,dsfont}
\usepackage{stackengine}
\usepackage{relsize}
\usepackage{comment}
\usepackage{physics}
\usepackage{braket}
\usepackage{xcolor}
\usepackage[T1]{fontenc}

\newif\ifhyper
\hypertrue
\ifhyper
\hypersetup{
   citecolor = {red},
   colorlinks = {true}, % false
   linkcolor = {blue},
   urlcolor = {blue} % magenta
}
\fi

\def\be{\begin{equation}}
\def\ee{\end{equation}}
\def\bea{\begin{eqnarray}}
\def\eea{\end{eqnarray}}

\definecolor{green}{HTML}{76c893}

\begin{document}

\title{Quantum-critical and dynamical properties of the XXZ bilayer with long-range interactions}
\author{Patrick Adelhardt}
\affiliation{Friedrich-Alexander-Universit\"at Erlangen-N\"urnberg (FAU), Department Physik, Staudtstra{\ss}e 7, D-91058 Erlangen, Germany}
\affiliation{Joint Quantum Institute and Joint Center for Quantum Information and Computer Science, University of Maryland Department of Physics and National Institute of Standards and Technology, College Park, Maryland 20742, USA}
\author{Antonia Duft}
\affiliation{Friedrich-Alexander-Universit\"at Erlangen-N\"urnberg (FAU), Department Physik, Staudtstra{\ss}e 7, D-91058 Erlangen, Germany}
\author{Kai Phillip Schmidt}
\affiliation{Friedrich-Alexander-Universit\"at Erlangen-N\"urnberg (FAU), Department Physik, Staudtstra{\ss}e 7, D-91058 Erlangen, Germany}

\begin{abstract}
		We study the XXZ square lattice bilayer model with antiferromagnetic non-frustrating long-range interactions that decay as a power law with the distance. Employing large-scale high-order series expansions with classical Monte Carlo integration (pCUT+MC) about the limit of isolated Heisenberg dimers in the rung-singlet phase, we investigate the one-triplon dispersion and the corresponding spectral weight along the parameter axes of the long-range decay exponent and the XXZ anisotropy. By tuning the latter, we observe two extended regions of 3d XY and Ising universality as well as 3d Heisenberg critical exponents at the isotropic point. Along the decay exponent axis, we demonstrate mean-field behavior for strong long-range couplings, the aforementioned three universality classes for sufficiently weak interactions, and continuously varying critical exponents in-between. Using extrapolations we are able to determine the one-triplon dispersion in a quantitative fashion up to the quantum-critical breakdown of the rung-singlet phase. This allows to extract the dynamical critical exponent $z$ as a function of the decay exponent, displaying a universal behavior. The detected $z<1$ for small decay exponents is in agreement with the expected properties of the anomalous Goldstone modes in the ordered phases with broken continuous symmetry.	
\end{abstract}

\maketitle

% Introduction
%%%%%%%%%%%%%%%%%%%%%%%%%%%%%%%%%%%%%%%%%%%%%%%%%%%%%%%%%%%%%%%%%%%%%%%%%%%%%%%%%%%%%%%%%%%%
\section{Introduction}

Correlated quantum many-body systems are fascinating because already very simple Hamiltonians can give rise to a plethora of intriguing quantum phases of matter featuring exotic ground states and novel collective behavior, such as quantum spin liquids, superconductivity, topological order, or fractional quasiparticle excitations with anyonic particle statistics. 
Usually, in condensed matter physics, correlations are induced by local, short-ranged interactions due to screening effects in the material and it is often justified to consider nearest-neighbor interactions as effective descriptions. 
However, in some physical systems it is necessary to consider the full algebraically decaying long-range interactions $\sim r^{-(d+\sigma)}$ with $d$ the dimension and $\sigma$ the decay exponent. 
This includes condensed matter systems \cite{Bitko1996,Chen2023,Ronnow2005,Burch2018,Gingras2011,Wang2022,Tiwari2024} like exotic spin ice materials on the pyrochlore lattice \cite{Harris1997, Ramirez1999,Bramwell2001,Castelnovo2008} as well as modern atomic, molecular, and optical (AMO) platforms \cite{Douglas2015,Vaidya2018,Defenu2023,Jaksch1998,Greiner2002,Bloch2005,Bloch2008,Baier2016,Gross2017,Schaefer2020,Su2023, Barredo2015,Barredo2016,Labuhn2016,Lienhard2018,Leseleuc2019,Browaeys2020,Scholl2021,Samajdar2021,Semeghini2021, Britton2012,Islam2013,Jurcevic2014,Richerme2014,Monroe2021}. The interest in quantum systems with long-range interactions has therefore been growing steadily over recent years. Due to the advancements in the control of individual entities it is now possible to realize a mesoscopic number of particles on different lattice geometries. 
Prominent platforms are trapped neutral atoms in cavities \cite{Douglas2015,Vaidya2018,Defenu2023}, ultracold dipolar atomic and molecular gases in optical lattices \cite{Jaksch1998,Greiner2002,Bloch2005,Bloch2008,Baier2016,Gross2017,Schaefer2020,Su2023}, Rydberg atom quantum simulators \cite{Barredo2015,Barredo2016,Labuhn2016,Lienhard2018,Leseleuc2019,Browaeys2020,Scholl2021,Samajdar2021,Semeghini2021,Defenu2023}, and trapped ion systems \cite{Britton2012, Islam2013, Jurcevic2014, Richerme2014, Monroe2021,Defenu2023}. While the long-range decay exponent $\sigma$ is fixed for most systems, for trapped ions it is even possible to continuously tune the decay exponent. 

On the fundamental side, long-range interacting systems can exhibit intrinsically different and exotic properties compared to their short-range cousins. 
They can give rise to Devil's staircase ground-state phase diagrams \cite{Bak1982, Koziol2023, Koziol2024} and unconventional critical properties like continuously varying critical exponents upon tuning the decay exponent $\sigma$ \cite{Fisher1972, Sak1973, Sak1977, Dutta2001, Defenu2015, Defenu2017, Defenu2020, Fey2016, Zhu2018, Fey2019, Adelhardt2020, Koziol2021, Langheld2022, Adelhardt2023, Shiratani2023, Song2023, Zhao2023}. Further, they can alter the dynamic low-energy spectral properties \cite{Yusuf2004, Laflorencie2005, Frerot2017, Diessel2023, Song2023b}, lead to the violation of area-law scaling of the entanglement entropy and to new scaling behavior \cite{Koffel2012, Vodola2014, Vodola2016, Frerot2017, Li2021, Zhao2024}, and cause the breakdown of the notion of causality due to the absence of Lieb-Robinson bounds \cite{Eisert2013, Hauke2013, Frerot2018, Vanderstraeten2018}. 
Often, the long-range transverse-field Ising model (LRTFIM), which possesses a discrete symmetry, served as a paradigmatic model for numerical techniques to study the quantum-critical behavior of long-range quantum systems \cite{Fey2016, Adelhardt2020, Koziol2021, Langheld2022, Zhu2018, Puebla2019, Shiratani2023, Fey2019, Koziol2021, Koffel2012, Sun2017, Koziol2019, Adelhardt2024} probing field-theoretical predictions from the long-range $O(n)$ quantum rotor model \cite{Fisher1972, Sak1973, Sak1977, Dutta2001, Defenu2015, Defenu2017, Defenu2020} and to study the intriguing interplay of long-range interactions and frustration \cite{Koffel2012, Fey2016, Sun2017, Fey2019, Koziol2019, Adelhardt2024, Duft2024}.
Altogether, long-range interactions open a different and exciting path in the study of correlated quantum many-body systems.

Recent studies \cite{Adelhardt2023, Song2023, Zhao2023, Diessel2023, Song2023b} went beyond the LRTFIM and investigated not only the critical behavior of long-range Heisenberg models with continuous $SU(2)$ symmetry on different lattice geometries as a function of the decay exponent $\sigma$, but also extracted dynamical low-energy properties in the ordered phase \cite{Diessel2023, Song2023b}. 
Interestingly, long-range interacting quantum many-body system with continuous symmetry can be relieved from fundamental paradigms of statistical physics, for example, through opening a gap via a generalized Higgs mechanism \cite{Diessel2023}, circumventing Goldstone's theorem \cite{Nambu1960, Goldstone1961, Goldstone1962}, or through continuous symmetry breaking circumventing the Hohenberg-Mermin-Wagner theorem \cite{Mermin1966a, Mermin1966b, Hohenberg1967, Coleman1973, Pitaevskii1991} as shown in theoretical studies \cite{Laflorencie2005, Kumar2013, Li2015, Tang2015, Gong2016a, Gong2016b, Maghrebi2017, Frerot2017, Ren2020, Yang2021, Yang2022} and confirmed in experimental setups \cite{Feng2023, Chen2023, Tiwari2024}. 
Even beyond, it was shown that deconfined quantum critical phase transitions become possible in one dimension \cite{Yang2020, Romen2024}. 
In this work, we continue the theoretical endeavor and study the antiferromagnetic XXZ square lattice bilayer model with non-frustrating (staggered) long-range interactions along the layers. 
To this end we employ large-scale high-order series expansions with classical Monte Carlo integration (pCUT+MC) \cite{Fey2019,Adelhardt2024} about the limit of isolated Heisenberg dimers in the rung-singlet phase. 
This allows to extract the rich quantum-critical properties and infer the dynamical low-energy properties of the ordered phase from studying the critical breakdown.  

The paper is organized as follows. 
We introduce the model in Sec.~\ref{sec:model} and give a short overview over the methodological aspects of this work in Sec.~\ref{sec:approach}. 
In Sec.~\ref{sec:results} we present the results starting with Subsec.~\ref{subsec:lr_axis} discussing the critical point and exponents as a function of the decay exponent $\sigma$, which is followed by an analog discussion of the same quantities along the XXZ anisotropy parameter $\theta$ in Subsec.~\ref{subsec:aniso_axis}. 
We also examine the low-energy dispersion in the rung-singlet phase up to the quantum-critical point, which allows us to extract dynamical low-energy properties in Subsec.~\ref{subsec:dyn_properties}. 
Finally, we summarize our results and draw conclusions in Sec.~\ref{sec:conclusions}.

% Model
%%%%%%%%%%%%%%%%%%%%%%%%%%%%%%%%%%%%%%%%%%%%%%%%%%%%%%%%%%%%%%%%%%%%%%%%%%%%%%%%%%%%%%%%%%%%
\section{Model}
\label{sec:model}

\begin{figure}[t]
	\centering
	\includegraphics[width=1.\columnwidth]{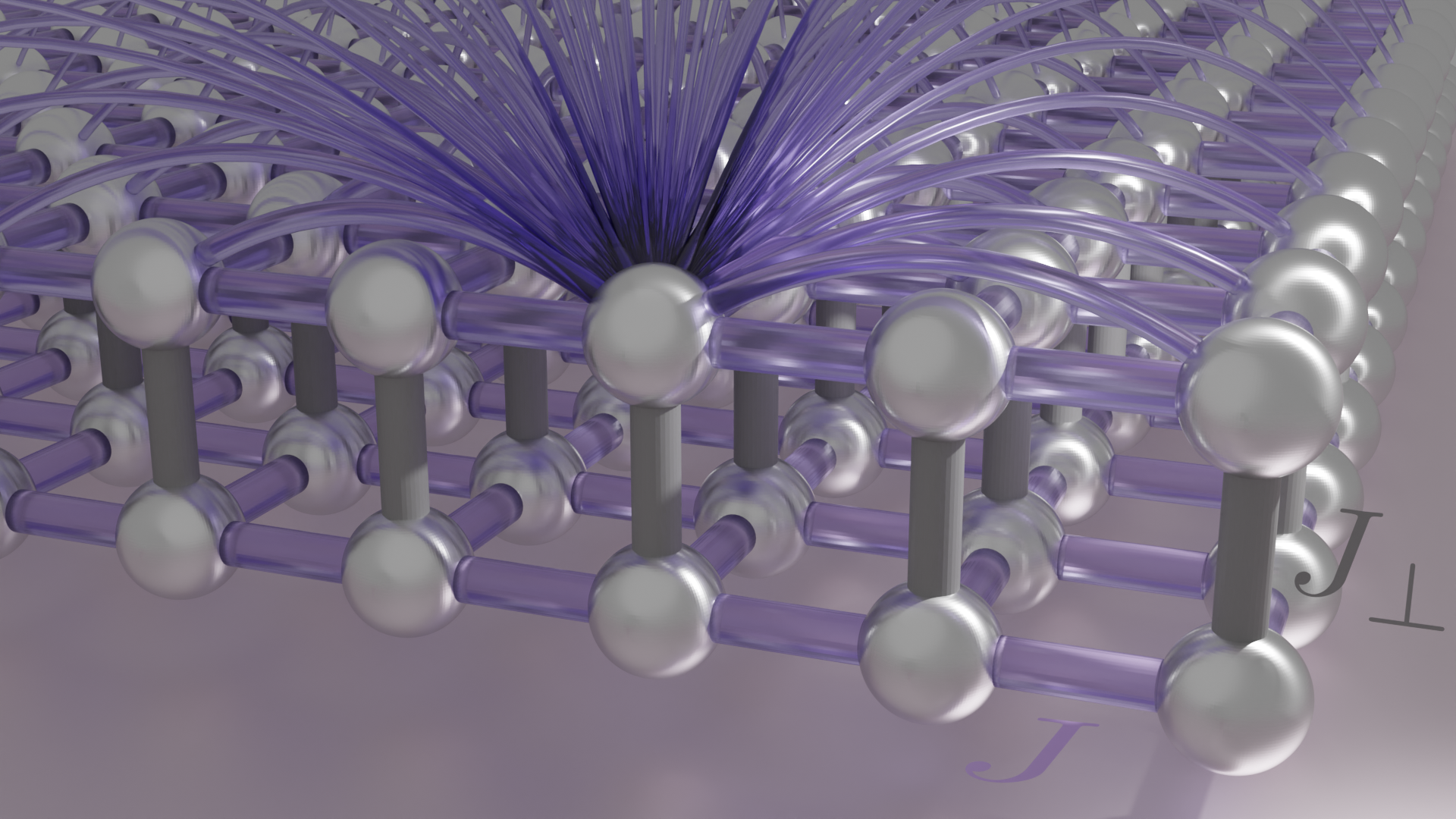}
	\caption{Illustration of the square lattice bilayer model with long-range interactions. Rung dimers represent the elementary unit cell. The antiferromagnetic Heisenberg interaction with coupling strength $J_{\perp}$ in each rung dimer is illustrated as a gray bond. 
	    Nearest-neighbor intralayer XXZ interactions with coupling strength $J$ are depicted by purple bonds. The long-range XXZ interactions beyond the nearest-neighbors are illustrated exemplary for a single site by purple curved bonds.}
	\label{fig:fig1}
\end{figure}

We investigate an antiferromagnetic square lattice bilayer model with non-frustrating (staggered) XXZ long-range interactions along the layers and nearest-neighbor Heisenberg couplings forming rung dimers connecting the two layers (see Fig.~\ref{fig:fig1} for illustration).
%%%%%%%%%%%%%%%%%%%%%%%%%%%%%%%%%%%%%%%%%%%%%%%%%%%%%%%%%%%%%%%%%%%%%%%%%%%%%%%%%%%%%%%%%%%%
%%%%%%%%%%%%%%%%%%%%%%%%%%%%%%%%%%%%%%%%%%%%%%%%%%%%%%%%%%%%%%%%%%%%%%%%%%%%%%%%%%%%%%%%%%%%
The Hamiltonian is given as 
\begin{equation}
	\mathcal{H} = J_{\perp} \sum_{\bm{i}} \vec{S}_{\bm{i}, 1} \vec{S}_{\bm{i}, 2} - \frac{1}{2}\sum_{\bm{i}\neq \bm{j}} \sum_{n = 1}^{2} J(\bm{j}-\bm{i}) 
	\mathcal{V}_{\bm{i},\bm{j},n}^{\rm XXZ}(\theta)\,,
	%\vec{S}_{\bm{i}, n}\vec{S}_{\bm{j}, n}
	\label{eq:ham_bilayer}
\end{equation}
with
\begin{equation}
 \mathcal{V}_{\bm{i},\bm{j},n}^{\rm XXZ}(\theta)= \cos\theta\,\left(S^x_{\bm{i}, n}S^x_{\bm{j}, n} + S^y_{\bm{i}, n}S^y_{\bm{j}, n}\right) + \sin\theta\,S^z_{\bm{i}, n}S^z_{\bm{j}, n}.
\label{eq:anisotropic_int}
\end{equation}
Here $\bm{i}$ denotes the rung position and $n$ refers to the layer $n\in\{1,2\}$.
The spin components are defined as $S_{\bm{i},n}^{\kappa} = \sigma_{\bm{i},n}^{\kappa}/2$ where $\sigma^{\kappa}$ with $\kappa\in\{x,y,z\}$ are Pauli matrices. 
The first term in Eq.~\eqref{eq:ham_bilayer} couples spins along the rungs with coupling strength $J_{\perp}>0$ and the second term describes the XXZ intralayer interaction with the coupling strength $J(\bm{\delta})$ with $\bm{\delta}=\bm{j}-\bm{i}$.
The XXZ anisotropy is parameterized by the angle $\theta$.
In the limit $\theta=0$ we have XY interactions only, for $\theta=\pi/4$ isotropic Heisenberg interactions, and in the limit $\theta=\pi/2$ Ising interactions. 
In the intermediate regime $0<\theta<\pi/4$ the XY interactions are dominant while in the other intermediate regime $\pi/4<\theta<\pi/2$ Ising interactions dominate.
 
The nearest-neighbor analog of the model has been subject to thorough scrutiny for some time \cite{Hida1990,Hida1992,Gelfand1996b,Zheng1997, Sandvik1994, Sandvik1995,Wang2006, Wenzel2008,Collins2008, Lohoefer2015}. 
Here, we consider couplings $J(\bm{\delta})$ that go beyond nearest-neighbor interactions following the power-law behavior 
\begin{equation}
	J(\bm{\delta}) = J\frac{(-1)^{\norm{\bm{\delta}}_1}}{|\bm{\delta}|^{2+\sigma}}
	\label{eq:stag_lr_int}
\end{equation}
with the strength of the interactions determined by the linear coupling constant $J>0$ and the power-law decay exponent $2+\sigma$. 
In the limiting case of $\sigma=\infty$ we recover antiferromagnetic nearest-neighbor interactions while in the other limit $\sigma=-2$ we have up to a sign uniform all-to-all coupling. 
We introduce an alternating sign factor $(-1)^{\norm{\bm{\delta}}_1}$ to ensure there is no frustration. 
In our notation, we distinguish between the one-norm $\norm{\cdot}_1$ in the exponent of the numerator to account for the correct sign of the interaction and the usual two-norm $|\cdot|\equiv\norm{\cdot}_2$ in the denominator to measure the real-space distance between two interacting spins. 
 
In the remainder of this article we will examine the quantum-critical properties of this model as a function of the coupling ratio $J/J_{\perp}$, decay exponent $\sigma$, and anisotropy parameter $\theta$. To this end we will perform high-order series expansions about the limit of isolated rung dimers $J=0$. 
We therefore introduce the perturbation parameter $\lambda=J/J_{\perp}$ by rescaling the energy spectrum of $\mathcal{H}$ and identifying the left term in Eq.~\eqref{eq:ham_bilayer} as the unperturbed part $\mathcal{H}_0$ and the right one as the perturbation $\mathcal{V}$. 
The unperturbed part $\mathcal{H}_0$ describing isolated rung dimers can be readily diagonalized.
The ground state of a single dimer is given by the lowest lying singlet state with total spin $S=0$
\begin{equation}
	\ket{s} = \frac{1}{\sqrt{2}}\left(\ket{\uparrow\downarrow}-\ket{\downarrow\uparrow}\right)
\end{equation} 
associated with the energy $\epsilon_0=-3/4$ and triplet excitations with total spin $S=1$ are given by
\begin{align}
	\ket{t^-} &= \ket{\downarrow\downarrow}, \\
	\ket{t^0} &= \frac{1}{\sqrt{2}}\left(\ket{\uparrow\downarrow}+\ket{\downarrow\uparrow}\right), \\
	\ket{t^+} &= \ket{\uparrow\uparrow},
\end{align}
with eigenenergy $1/4$.
Hence, the ground state of the unperturbed Hamiltonian $\mathcal{H}_0$ is given by the product state of rung-singlet states $\ket{s}$ with triplets $\ket{t^{\alpha}_i}$ at an arbitrary rung $i$ and $\alpha\in\{-,0,+\}$ as elementary excitations above the ground state. 
We then express the Hamiltonian in Eq.~\eqref{eq:ham_bilayer} exactly by hard-core bosonic operators $t_{\bm{i},\alpha}^{(\dagger)}$ annihilating (creating) triplet excitations at site $\bm{i}$, which yields an Hamiltonian in second quantization on an effective square lattice formed by the rung dimers (see App.~\ref{app:processes} listing all quasiparticle processes of the Hamiltonian). 
For small but finite $\lambda$ the ground state is adiabatically connected to the trivial product state of rung singlets and elementary triplet excitations become dressed by quantum fluctuation giving rise to an effective quasiparticle referred to as triplon \cite{Schmidt2003}. 

Upon further increasing the perturbation parameter $\lambda$ we expect a second-order quantum phase transition towards an antiferromagnetic ordered phase which spontaneously breaks the symmetry of the Hamiltonian. 
For $0\le\theta\le\pi/4$ this is associated with the breaking of continuous symmetries. 
At the isotropic Heisenberg point $\theta=\pi/4$ the ground state breaks the $SU(2)$ symmetry of the Hamiltonian and for $0\le\theta<\pi/4$ the $U(1)$ symmetry in the XY plane. 
For $\pi/4<\theta\le\pi/2$ the ground states breaks the $\mathbb{Z}_2$ symmetry of $S^z$ spins. 
Due to Goldstone's theorem \cite{Nambu1960,Goldstone1961, Goldstone1962}, which states that continuous symmetry breaking is accompanied by the presence of massless Nambu-Goldstone modes in the ordered phase, we expect the respective antiferromagnetic phases with broken continuous symmetry to feature massless magnon excitations, while the antiferromagnetic phase due to $\mathbb{Z}_2$-symmetry breaking is gapped as known from the transverse-field Ising model (TFIM) \cite{Pfeuty1970, Pfeuty1976}.

% Methods
%%%%%%%%%%%%%%%%%%%%%%%%%%%%%%%%%%%%%%%%%%%%%%%%%%%%%%%%%%%%%%%%%%%%%%%%%%%%%%%%%%%%%%%%%%%%
\section{Approach}
\label{sec:approach}

In this section we provide a brief overview of the methodological aspects of high-order series expansions for quantum systems with long-range interactions. 
This approach was pioneered in Ref.~\cite{Fey2019} combining the well-established method of perturbative continuous unitary transformations (pCUT) \cite{Knetter2000, Knetter2003} with classical Monte Carlo integration (MC) specifically to tackle long-range interacting quantum spin systems \cite{Fey2019}. 
We refer to a recent comprehensive introduction to the pCUT+MC method in Ref.~\cite{Adelhardt2024} for the interested reader.
Here, we try to condense the approach to the main ideas to be self-contained. 
Note that this section is not essential for the discussion of the results in Sec.~\ref{sec:results}.

\subsection{Perturbative continuous unitary transformations}

The Hamiltonian in Eq.~\eqref{eq:ham_bilayer} can be interpreted as the usual perturbation problem of the form
\begin{equation}
	\mathcal{H} = \mathcal{H}_0 + \lambda\mathcal{V},
\end{equation}
where the unperturbed part $\mathcal{H}_0$ describes isolated rung dimers and the perturbation $\mathcal{V}$ associated with the perturbation parameter $\lambda=J/J_{\perp}$ describes the long-range interactions along the layers. 
Here, the spectrum of the unperturbed part is equidistant and bounded from below and we can express this part through the counting operator $Q=\sum_{\bm{i},\alpha}  t_{\bm{i},\alpha}^{\dagger}t_{\bm{i},\alpha}^{\phantom{\dagger}}$ counting the number of triplet quasiparticles (qps). 
Moreover, we can write the perturbation as
\begin{equation}
	\mathcal{V} = T_{-2} + T_0 + T_{+2}
\end{equation}
in terms of $T_n$-operators that contain all processes changing the quasiparticle number by $n$ quanta. We can see in Tab.~\ref{tab:ham_processes} in App.~\ref{app:processes} that the Hamiltonian indeed only contains processes that do not change the particle number ($T_0$) or that change the number of quasiparticles exactly by two ($T_{\pm 2}$). 
Because of the properties above, the method of perturbative continuous unitary transformations (pCUT) can be applied to our system \cite{Knetter2000}. 

The idea of the pCUT method is to transform the original Hamiltonian, order by order in the perturbation parameter, to an effective quasiparticle-conserving Hamiltonian $\mathcal{H}_{\text{eff}}$ finding an optimal basis in which the many-body problem reduces to an effective few-body problem. 
To this end, we introduce a unitary transformation $H(\ell) = U^{\dagger}(\ell)\mathcal{H}U(\ell)$ depending on a continuous parameter $\ell$. 
We require that the transformation recovers the original perturbative problem for $\ell=0$ and that it maps to the desired effective description in the limit $\ell\rightarrow\infty$. 
We can write the effective Hamiltonian in a generic form given by
\begin{equation}
	\mathcal{H}_{\text{eff}} = \mathcal{H}_0 + \hspace{-0.3cm}\sum_{\substack{k=1 \\ \sum_j n_j = k}}^{\infty} \hspace{-0.2cm}\lambda_{1}^{n_1} \dots \lambda_{N_\lambda}^{n_{N_\lambda}} \hspace{-0.3cm} \sum_{\substack{\dim(\bm{m})=k, \\ \sum_i m_i=0}} \hspace{-0.3cm}\mathcal{C}(\bm{m})\,T_{m_1}\dots T_{m_k},
	\label{eq:eff_ham_pcut}
\end{equation}
where we introduced $N_{\lambda}$-many perturbation parameters $\lambda_i$ and the condition $\sum_i m_i=0$ constraining the product of $T_{m_i}$ to be quasiparticle conserving \cite{Knetter2000}. 
The coefficients $C(\bm{m})\in\mathbb{Q}$ are exact and rational. However, the effective Hamiltonian $\mathcal{H}_{\text{eff}}$ is not normal-ordered which leads to the second, model dependent step of a pCUT calculation the normal ordering.

Here, we concentrate on the 1qp block of $\mathcal{H}_{\text{eff}}$ which corresponds after normal-ordering of $\mathcal{H}_{\text{eff}}$ for each triplon flavor $\alpha\in\{-,0,+\}$ to a one-particle hopping problem on the square lattice. Exploiting the translation symmetry and introducing triplet creation and annihilation operators with momentum $\bm{k}$  by \mbox{$t_{\bm{k},\alpha}^{\dagger}=(2/\sqrt{N})\sum_{\bm{i}}\exp({\rm i}\bm{k}\cdot \bm{i})\, t_{\bm{i},\alpha}^{\dagger}$} with $N$ the number of spins, the effective 1qp Hamiltonian can be diagonalized 
\begin{equation}
	\mathcal{H}_{\text{eff}}^{\rm 1qp} = \bar{E}_0 + \sum_{\bm{k}} \omega^{\text{1qp}}_{\alpha}(\bm{k})\, t_{\bm{k},\alpha}^{\dagger} t_{\bm{k},\alpha}^{\phantom{\dagger}}\, .
\end{equation}
Here, $\bar{E}_0$ is the dressed ground-state energy and $\omega^{\text{1qp}}_{\alpha}(\bm{k})$ is the one-triplon dispersion for each flavor $\alpha$. Below we will calculate this dispersion as a high-order series in the perturbation parameter $\lambda$.

The pCUT method does not only allow to determine the effective Hamiltonian but also allows for the calculation of effective observables \cite{Knetter2003}. 
It is possible to derive a general expression of an effective observable in analogy to Eq.~\eqref{eq:eff_ham_pcut}, which is given by 
\begin{equation}
\begin{split}
	\mathcal{O}_{\text{eff}} &= \sum_{\substack{k=1 \\ \sum_j n_j = k}}^{\infty} \lambda_{1}^{n_1} \dots \lambda_{N_{\lambda}}^{n_{N_{\lambda}}} \\
&\times \sum_{i=1}^{k+1}\sum_{\dim(\bm{m})=k} \hspace{-0.2cm}\tilde{\mathcal{C}}(\bm{m}; i)T_{m_1}\dots T_{m_{i-1}}\mathcal{O} T_{m_i}\dots T_{m_{k}},
\end{split}
\label{eq:eff_obs_pcut}
\end{equation}
where $\mathcal{O}$ is a generic observable of interest and the coefficients $\tilde{\mathcal{C}}(\bm{m};i)$ are exact rational coefficients \cite{Knetter2003}. 
In contrast to $\mathcal{H}_{\text{eff}}$ the effective observable is not quasiparticle conserving as it contains no such restriction. 
Here, we want to calculate spectral weights which are defined via the dynamic structure factor
\begin{equation}
	\mathcal{S}_{\alpha}(\bm{k}, \omega) = \frac{1}{2\pi N} \sum_{\bm{i}, \bm{j}} \int_{-\infty}^{\infty} \hspace{-0.3cm} \mathrm{d}t \, e^{\text{i}[\omega t - \bm{k}(\bm{j} - \bm{i})]} \langle \mathcal{O}_{\bm{j}}^{\alpha}(t)\mathcal{O}_{\bm{i}}^{\alpha}(0)\rangle\ ,
\end{equation}
that is the real-space and time Fourier transform of the correlation function $\langle \mathcal{O}_{\bm{j}}^{\alpha}(t)\mathcal{O}_{\bm{i}}^{\alpha}(0)\rangle$ of an observable $\mathcal{O}^{\alpha}$. 
The dynamic structure factor can be decomposed into the individual contributions of its energy eigenstates $\ket{\psi_{\Lambda}}$ with $\Lambda$ denoting a set of quantum numbers as follows
\begin{equation}
	\mathcal{S}_{\alpha}(\bm{k}, \omega) = \sum_{\Lambda} \mathcal{S}^{\Lambda}_{\alpha}(\bm{k}, \omega) .
\end{equation}
We restrict to the set of quantum numbers $\Lambda\equiv 1\text{qp}$ for the relevant 1qp contributions and by integrating out the time dependence (see Refs.~\cite{Hamer2003, Adelhardt2024}), we arrive at 
\begin{equation}
	\mathcal{S}^{\text{1qp}}_{\alpha}(\bm{k},\omega) = \delta(\omega-E_{\text{1qp}}+E_0)\,\mathcal{S}^{\text{1qp}}_{\alpha}(\bm{k}), 
	\label{eq:1qp_sw}
\end{equation}
where $\mathcal{S}^{\text{1qp}}_{\alpha}(\bm{k})$ is the 1qp spectral weight and $E_{\text{1qp}}$ the 1qp band.
We determine the spectral weight, choosing the antisymmetric spin combinations
\begin{align}
	\mathcal{O}^{x}_i &=  \frac{1}{\sqrt{2}}\left(S^x_{i,1}-S^x_{i,2}\right)\, \\
	\mathcal{O}^{z}_i &=  \frac{1}{2}\left(S_{i,1}^z-S_{i,2}^z\right)
	\label{eq:spin_obs}
\end{align}
as the observables  $\mathcal{O}^{\alpha}_i$. 
The observables expressed as processes in the triplet basis are given in Tab.~\ref{tab:obs_processes} in App.~\ref{app:processes}. 
This choice for the observables is not arbitrary since in the bilayer model the antisymmetric combination of spin operators couples only to odd quasiparticle numbers while a symmetric choice would couple only to even quasiparticle numbers \cite{Knetter2003b}. 
This exact property originates from the exact reflection symmetry of the bilayer about the center line.
The choice of different prefactors comes from the fact that $\mathcal{O}_i^{z}$ couples to a single triplon flavor, while $\mathcal{O}_i^{x}$ couples to two triplon flavors mixing the contributions of $\mathcal{O}_i^{x}$ and $\mathcal{O}_i^{y}$. %$=-i/\sqrt{2}\,(S^y_{i,1}-S^y_{i,2})$.
Again, the goal is to calculate the normal-ordered effective observable and the 1qp spectral weight $\mathcal{S}^{\text{1qp}}_{\alpha}(\bm{k})$ as a high-order series in the perturbation parameter $\lambda$. 

\subsection{Monte Carlo Embedding}

The effective pCUT Hamiltonian given in Eq.~\eqref{eq:eff_ham_pcut} as well as effective pCUT observable in Eq.~\eqref{eq:eff_obs_pcut} are generic expressions and model independent coming at the cost of an additional normal ordering that we need to perform. 
This is usually done by applying $\mathcal{H}_{\text{eff}}$ to finite clusters. 
Most efficiently, this is done by a linked-cluster expansion implemented as a full graph decomposition exploiting the linked-cluster theorem \cite{Coester2015}. 
For long-range interacting systems a conventional graph decomposition would lead to infinitely many graph contributions already at first order of perturbation \cite{Fey2016}. 
Therefore, it is necessary to use white graphs \cite{Coester2015,Fey2016} where we associate an abstract perturbation parameter $\lambda_i$ with each edge of a graph and it is only during the embedding of the graph contribution on the actual lattice that the abstract perturbation parameter is substituted with the actual algebraically decaying long-range coupling strength \cite{Fey2019, Adelhardt2024}.

Embedding the white-graph contributions for long-range interactions is different to its short-range counterpart because quasiparticle processes are not restricted to a finite range given by the order of the process but can cover the entire infinite lattice. 
The long-range embedding problem constitutes $n_s$-many infinite sums over the graph contributions $f_{n_s, \bm{k}}^{(\mathfrak{o})}$ with $n_s$ being the number of sites and $\mathfrak{o}$ the perturbative order of the contribution. 
The embedding sum $S[\cdot]$ can be expressed as 
\begin{equation}
	S[f_{n_s, \bm{k}^{\star}}^{(\mathfrak{o})}] = \sum_{\bm{i}_1}{\vphantom{\sum}}'\dots\sum_{\bm{i}_{n_s}}{\vphantom{\sum}}' f_{n_s, \bm{k}^{\star}}^{(\mathfrak{o})}(\bm{i}_1,\dots, \bm{i}_{n_s})\ ,
	\label{eq:mc_sum_1qp}
\end{equation}
where we evaluate the contributions at a specific momentum $\bm{k}^{\star}$ and the primed sums over the indices $\bm{i}_j$ are a short notation for excluding overlaps between the positions of graph vertices on the lattice \footnote{The $k$-dependence comes due to  cosine terms that appear as we Fourier transform the 1qp particle processes}. 
This is a high-dimensional integration problem of infinite nested sums that is extremely challenging to solve with conventional integration technique \cite{Fey2016}. 
This is why we use Monte Carlo (MC) integration which is much more suitable for such problems \cite{Fey2019, Adelhardt2024}. 
We use an MC algorithm that essentially consists of two MC moves. 
In both of these moves we choose a random site and change the current embedding by moving the sites by randomly drawing new positions. 
Depending on how we draw the new positions, we have one move that is intended to induce small fluctuations and another one that is there to capture the correct asymptotic behavior of the algebraically decaying long-range interactions. 
In the hard-core bosonic triplet formulation introduced above the rung dimers can be naturally understood as supersites, thus the embedding must be performed on a simple square lattice.

We calculate the 1qp dispersion  $\omega^{\text{1qp}}_{\alpha}(\bm{k})$ as well as $s^{\text{1qp}}_{\alpha}(\bm{k})$ which is defined by the 1qp spectral weight $\mathcal{S}^{1\text{qp}}_{\alpha}(\bm{k}) \equiv |s^{\text{1qp}}_{\alpha}(\bm{k})|^2$ as a high-order series expansion given by
\begin{equation}
	\kappa(\bm{k}^{\star}) = p_0 + \sum_{\mathfrak{o}=1}^{\mathfrak{o}_{\text{max}}} p_{\mathfrak{o}} \lambda^{\mathfrak{o}} \quad\text{with}\quad p_{i>0} = \sum_{n_s=2}^{\mathfrak{o}+1} \bar{S}[f_{n_s,\bm{k}^{\star}}^{(\mathfrak{o})}]\ ,
	\label{eq:mc_series}
\end{equation}
where $\kappa(\bm{k})\in \{\omega^{\text{1qp}}_{\alpha}(\bm{k}),s^{\text{1qp}}_{\alpha}(\bm{k})\}$. 
The bar over $S[\cdot]$ indicates that we average the MC sum over several seeds. 
We calculated the high-order series up to order 10 for the 1qp dispersion  $\omega^{\text{1qp}}(\bm{k})$ for several momenta $\bm{k}$ including the gap momentum $\bm{k}_c$ and up to order 9 for the contribution $s^{\text{1qp}}_{\alpha}(\bm{k})$ of the 1qp spectral weight. 
Note, for nearest-neighbor interactions ($\sigma=\infty$) we cannot use the MC algorithm so we used conventional embedding for the 1qp gap reaching order 11. 
Using an appropriately designed single cluster for the 1qp spectral weight we were able to determine the 1qp spectral weight up to order 8. 

\subsection{Extrapolation}
\label{subsec:extrapolation}

To extract the critical point and the associated critical exponent from the high-order series we employ DlogPadé extrapolations. 
The Padé approximant of a quantity $\kappa$ is defined as
\begin{equation}
	P[L, M]_{\kappa} = \frac{P_L(\lambda)}{Q_{M}(\lambda)} = \frac{p_0 + p_1\lambda + \cdots + p_L\lambda^L }{1 + q_1\lambda + \cdots + q_M\lambda^M}\ ,
	\label{eq:pade}
\end{equation}
where $p_i, q_i \in \mathbb{R}$ and the degrees $L$, $M$ of the numerator polynomial $P_{L}(\lambda)$ and denominator polynomial $Q_{M}(\lambda)$ are restricted to $\mathfrak{o}_{\text{max}} = L+M$, where $\mathfrak{o}_{\text{max}}$ is the maximal perturbative order. 
We can determine the coefficients $p_i$ and $q_i$ by equating the ansatz for the Padé approximant with the series given by Eq.~\eqref{eq:mc_series}. 
Along the same lines we can define the Padé approximant of the logarithmic derivative
\begin{equation}
	P[L, M]_{\mathcal{D}} = \dv{\lambda}\ln(\kappa)
	\label{eq:dlogpade}
\end{equation}
with $\mathfrak{o}_{\text{max}}-1=L+M$. 
The corresponding DlogPadé approximant can be obtained by integration. 
By analyzing the poles of the approximants and identifying the physical pole we can determine the critical point $\lambda_{\rm c}$ and by calculating the residuum $\vartheta=\Res P[L, M]_{\mathcal{D}}\vert_{\lambda=\lambda_{\rm c}}$ we can determine the associated critical exponent $\vartheta$.
If the critical point $\lambda_{\rm c}$ is already known we can use this $\lambda_{\rm c}$ to bias the approximant using
\begin{equation}
	P[L,M]_{\vartheta^{\star}} = (\lambda_{\rm c} - \lambda)\dv{\lambda}\ln(\kappa)\ .
	\label{eq:biaseddlogpade}
\end{equation}
The critical exponent $\vartheta$ can be determined by the residuum as well. 
To obtain reliable estimates for the critical quantities we arrange multiple DlogPadés \mbox{$L+M=\mathfrak{o}<\mathfrak{o}_{\text{max}}$} into families
with $L-M=\text{const.}$ and average over the highest-order approximants of each family. Additional information on the extrapolation and a discussion of the convergence behavior of the series obtained for the XXZ bilayer can be found in App.~\ref{app:convergence}.

% RESULTS
%%%%%%%%%%%%%%%%%%%%%%%%%%%%%%%%%%%%%%%%%%%%%%%%%%%%%%%%%%%%%%%%%%%%%%%%%%%%%%%%%%%%%%%%%%%%

\section{Results}
\label{sec:results}

In this section, we discuss the results obtained from the pCUT+MC approach. 
First, we examine the critical properties of the quantum phase transition from the trivial rung-singlet phase towards the symmetry-broken antiferromagnetic phases as a function of the long-range decay exponent $\sigma$. 
Second, we complement the discussion by examining the results along the anisotropy parameter $\theta$. 
Third, we show that we can reliably determine the dispersion at the critical point with which we can extract the dynamical critical exponent $z$ allowing us to make statements about the nature of gapless Goldstone modes inside the symmetry-broken phases.

% long-range axis
%%%%%%%%%%%%%%%%%%%%%%%%%%%%%%%%%%%%%%%%%%%%%%%%%%%%%%%%%%%%%%%%%%%%%%%%%%%%%%%%%%%%%%%%%%%%
\subsection{Long-range axis}
\label{subsec:lr_axis}

We set the anisotropy parameter $\theta$ to distinct values $\theta\in\{0,\pi/8,\pi/4,3\pi/8,\pi/2\}$ and analyze the critical point and critical exponents as a function of the long-range decay exponent $\sigma$. 
Depending on the anisotropy parameter $\theta$ there are different regimes of spontaneous symmetry breaking. 
For $0\le\theta<\pi/4$ the quantum phase transition occurs due to $U(1)$-symmetry breaking, at $\theta=\pi/4$ due to $SU(2)$-symmetry breaking, and in the regime $\pi/4<\theta\le\pi/2$ due to the breaking of the discrete $\mathbb{Z}_2$ symmetry of the Hamiltonian in Eqs.~\eqref{eq:ham_bilayer} and \eqref{eq:anisotropic_int}. 
We can understand the quantum phase transition in terms of the long-range $O(n)$ quantum rotor model with $n=1,2,3$ and we expect three distinct regimes of criticality as a function of $\sigma$. 
For $\sigma\ge 2-\eta_{\text{SR}}$ we expect the nearest-neighbor criticality of the respective regime, for $\sigma\le 4/3$ long-range mean-field behavior, and in-between a non-trivial regime of continuously varying critical exponents. 
Here, $\eta_{\text{SR}}$ is the anomalous dimension critical exponent of the respective short-range regime.

\begin{figure}[t]
	\centering
	\includegraphics[width=1.\columnwidth]{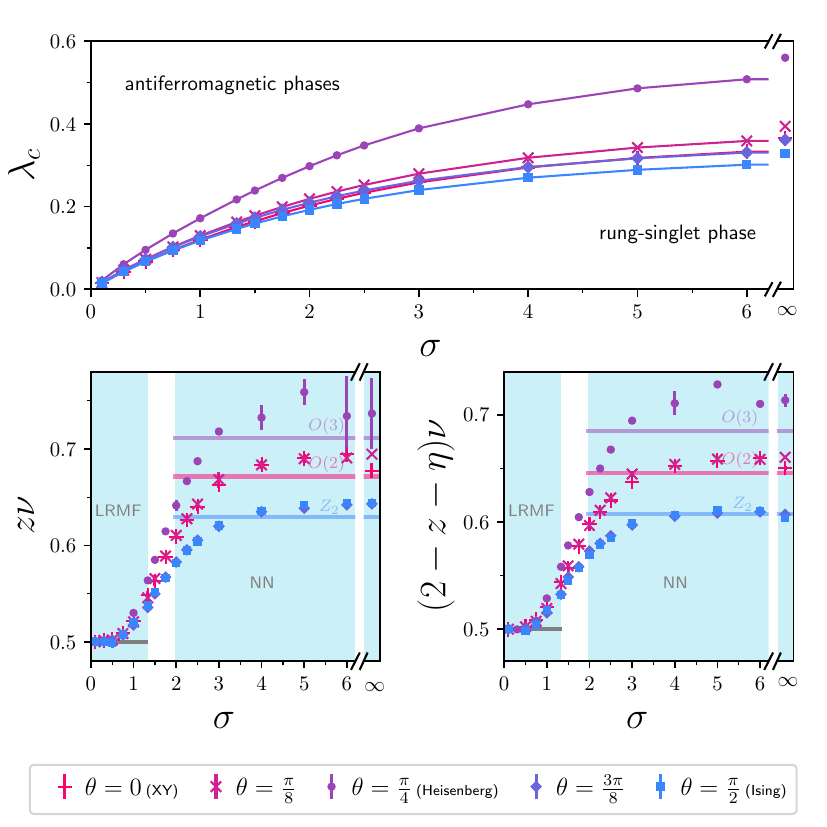}
	\caption{Critical point $\lambda_{\rm c}$ and critical exponents $z\nu$ and \mbox{$(2-z-\eta)\nu$} are shown as a function of the long-range decay exponent $\sigma$ including the limiting case $\sigma=\infty$ of nearest-neighbor (NN) interactions. 
The results are shown for distinct XXZ interactions with anisotropy parameter \mbox{$\theta\in\{0,\pi/8,\pi/4,3\pi/8,\pi/2\}$}. 
(\textit{Upper panel}): The critical line $\lambda_{\rm c}$ divides the rung-singlet from the antiferromagnetic phases. 
The rung-singlet phase is the most stable for isotropic Heisenberg interactions at $\theta=\pi/4$. 
(\textit{Lower panels}): The critical exponents $z\nu$ (left) and $(2-z-\eta)\nu$ (right) show three distinct critical regimes. 
The expected NN and long-range mean-field (LRMF) regimes are visualized by a blue background. 
The colorful solid lines in the NN regime depict the literature values \cite{Campostrini2001, Kos2016, Hasenbusch2011, Hasenbusch2019, Chester2021} for the critical exponents of the three expected universality classes associated with O(2), O(3), and $\mathbb{Z}_2$-symmetry breaking for XY, Heisenberg, and Ising interactions, respectively. 
The solid gray line shows the LRMF values of the critical exponent. 
The results for the critical exponents are in good agreement with the three critical regimes, although the boundaries of the non-trivial regime cannot be accurately resolved and the deviations in the NN regime are quite large for isotropic Heisenberg interactions.}
	\label{fig:fig2}
\end{figure}

We use high-order series expansions determined by the pCUT+MC approach for the 1qp gap \mbox{$\Delta^{\text{1qp}}=\omega_{\alpha}^{\text{1qp}}(\bm{k}_{\rm c})$}  to determine the critical point $\lambda_{\rm c}$ as a function of $\sigma$. 
Here, $\alpha$ is the respective critical band and $\bm{k}_{\rm c}=(\pi,\pi)$ the critical gap momentum due to the antiferromagnetic order in the symmetry-broken phases. 
In the basis above, the relevant triplon excitations with flavor $\alpha$, which condense at the quantum critical point, are the $\ket{t^{\pm}}$ triplons for $0\le \theta < \pi/4$ and $\ket{t^0}$ triplons for $\pi/4 < \theta \le \pi/2$. 
As the gap closes with a power-law behavior about the critical point
\begin{equation}
	\Delta^{\text{1qp}} \sim |\lambda-\lambda_{\rm c}|^{z\nu}
\end{equation}
we can also determine the critical gap exponent $z\nu$. 
Additionally, we use the critical point determined by the gap closing to determine a biased estimate for the critical exponent $(2-z-\eta)\nu$ from the power-law divergence 
\begin{equation}
	\mathcal{S}^{\text{1qp}}(\bm{k}_{\rm c}) \sim |\lambda-\lambda_{\rm c}|^{-(2-z-\eta)\nu}
\end{equation}
of the 1qp spectral weight. 
In Fig.~\ref{fig:fig2} we plot the critical point $\lambda_{\rm c}$ (upper panel) and the critical exponents $z\nu$ and \mbox{$(2-z-\eta)\nu$} (lower left and right panels, respectively) in dependence of the long-range decay exponent $\sigma$.
We observe that the antiferromagnetic phases are stabilized upon decreasing the decay exponent $\sigma$. 
For $\sigma\rightarrow 0$ the critical point shifts to zero. 
The antiferromagnetic phase is the largest for pure Ising interactions ($\theta=\pi/2$) and smallest for the isotropic Heisenberg point ($\theta=\pi/4$) as quantum fluctuations destroying the long-range order are enhanced in the Heisenberg model. 
At $\sigma=\infty$  with only nearest-neighbor Heisenberg interactions the critical value $\lambda_{\rm c}=0.5604(29)$ is in good agreement with the best literature values $\lambda_{\rm c}= 0.56075(23)$ \cite{Wang2006} and \mbox{$\lambda_{\rm c}= 0.56127(4)$} \cite{Wenzel2008} from QMC calculations.

In the lower part of Fig.~\ref{fig:fig2}, we see that the exponents determined with the pCUT+MC method approach three distinct constant values in the NN regime as expected from three different types of symmetry breaking. 
For small $\sigma$ in the LRMF regime the exponents go to their mean-field values. 
Deep enough in this regime for $\sigma\le 0.5$ the exponents are within $<1\%$ of the exact mean-field value. 
In the NN regime we overestimate the literature values notably, however, we can still clearly identify the three different NN universality classes. 
In the Ising limit the values are the closest while for Heisenberg interactions the deviation as well as the uncertainty from the extrapolations is the largest. 
The exact reason why the deviation is the largest for Heisenberg interactions is not obvious, but has been observed previously with series expansions for the NN case \cite{Coester2011Diploma}. 
The exponent $(2-z-\eta)\nu$ associated with the 1qp spectral weight shows smaller deviations than $z\nu$ which may be due to the fact that we used DlogPadé approximants biased with the critical point from the closing of the gap. 

In the intermediate regime $4/3<\sigma<2-\eta_{\text{SR}}$ between LRMF and NN criticality, the critical exponents smoothly interpolate between the two adjacent regimes as expected from field-theoretical considerations of the long-range $O(n)$ quantum rotor model  \cite{Fisher1972, Sak1973, Sak1977, Dutta2001, Defenu2015, Defenu2017, Defenu2020}. 
We find that the regime boundaries cannot be resolved well by the pCUT+MC approach but this is already known from previous studies \cite{Fey2019, Adelhardt2020, Adelhardt2023, Adelhardt2024}. 
At the upper critical dimension at $\sigma=4/3$ the presence of multiplicative logarithmic corrections to the dominant power-law behavior is well-known \cite{Wegner1973, Bauerschmidt2014} and can explain the deviation at the upper critical dimension. 
The presence of corrections to the power-law scaling at $\sigma=4/3$ in turn also influences the critical exponents in the vicinity. 
Interestingly, the deviation is largest for Heisenberg and smallest for Ising interactions which coincides with the expectations since the exponent of the logarithmic correction is larger the bigger the symmetry group \cite{Wegner1973, Bauerschmidt2014}. 

At the other interface between non-trivial and NN regime similar deviations can be observed. 
The reason for the deviation here is less clear. 
It could be due to the finite order of the series not being able to capture the abrupt change of the critical exponents or due to unknown corrections to the power-law behavior in this regime or a combination of both. 
We should point out however that analogous to deviations at the upper critical dimension the magnitude of the deviations also seems to depend on the symmetry group which hints towards the presence of unknown corrections.  
Altogether the results are in excellent agreement with all expectations. 
We can clearly identify the three critical regimes as a function of $\sigma$ as well as the three distinct NN universality regimes for different $\theta$.

% anisotropy axis
%%%%%%%%%%%%%%%%%%%%%%%%%%%%%%%%%%%%%%%%%%%%%%%%%%%%%%%%%%%%%%%%%%%%%%%%%%%%%%%%%%%%%%%%%%%%
\subsection{Anisotropy axis}
\label{subsec:aniso_axis}

\begin{figure}[t!]
	\centering
	\includegraphics[width=1.\columnwidth]{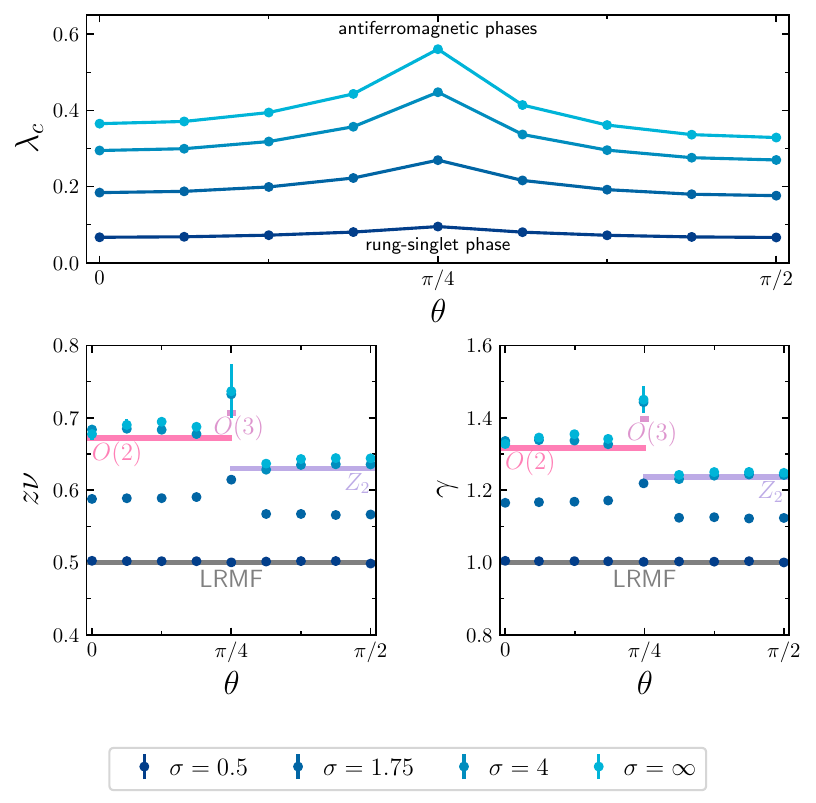}
	\caption{Critical point $\lambda_{\rm c}$ and exponents $z\nu$ and $\gamma$ as a function of the anisotropy parameter $\theta$. 
Results are shown for distinct long-range decay exponents $\sigma\in\{0.5,1.75,4,\infty\}$ in the three critical regimes. 
(\textit{Upper panel}): The critical lines $\lambda_{\rm c}$ for different $\sigma$ divide the rung-singlet phase from the antiferromagnetic phases. 
The rung-singlet phase is largest for $\sigma=\infty$ and smallest for $\sigma=0.5$. 
(\textit{Lower panel}): The critical exponents $z\nu$ (left) and $\gamma$ (right) are depicted. 
For $\sigma\in\{4,\infty\}$ the results are in line with $O(2)$-symmetry breaking for $0\le\theta<\pi/4$, $O(3)$-symmetry breaking at $\theta=\pi/4$, and $\mathbb{Z}_2$-symmetry breaking for $\pi/4<\theta\le\pi/2$. 
The expected values  \cite{Campostrini2001, Kos2016, Hasenbusch2011, Chester2021} are indicated by colorful lines. 
For $\sigma=0.5$, LRMF values (gray lines) are recovered for all $\theta$. 
In the non-trivial regime at $\sigma=1.75$ the three distinct regimes of symmetry breaking are still visible although the absolute values shift.}
	\label{fig:fig3}
\end{figure}

We now fix the long-range decay exponent to \mbox{$\sigma\in\{0.5,1.75,4,\infty\}$} and tune the anisotropy parameter $\theta$ from $\theta=0$ with pure XY interactions through the isotropic Heisenberg point at $\theta=\pi/4$ to pure Ising interactions at $\theta=\pi/2$. 
As in the previous section, we determine the critical point $\lambda_{\rm c}$ and the critical exponent $z\nu$. 
Further, we determine the critical exponent $\gamma$ as we can directly exploit the Fisher scaling relation $\gamma=(2-\eta)\nu$. 
The critical properties as a function of $\theta$ are shown in Fig~\ref{fig:fig3}. 
For the critical point $\lambda_{\rm c}$ we see a maximum at the Heisenberg point. 
Further, the critical points in the opposite limits $\theta=0$ and $\theta=\pi/2$ are not symmetric. 
Clearly, the extent of the antiferromagnetic phases becomes bigger as the long-range interaction becomes stronger by decreasing $\sigma$, which we already saw in Fig.~\ref{fig:fig2}. 

In the lower panels of Fig.~\ref{fig:fig3} we show the critical exponents $z\nu$ and $\gamma$. 
Again, we can clearly identify three regimes of universality in the NN regime in agreement with field-theoretical expectations. 
There are two extended regimes of XY criticality for $0\le \theta < \pi/4$ and Ising criticality for $\pi/4 < \theta \le \pi/2$ and a single point of Heisenberg criticality at $\pi/4$. 
In the non-trivial regime, where the critical exponents vary continuously, these three criticality regions as a function of $\theta$ are retained but the absolute values are lowered. 
Further tuning the long-range decay exponent $\sigma$ to smaller values and arriving at the LRMF regime, this behavior is changed as can be seen for $\sigma=0.5$, where we observe the expected mean-field exponents $z\nu=0.5$ and $\gamma=1$ for all $\theta$.

Overall, the critical properties along the $\theta$ anisotropy axis consolidate the previous assessments along the long-range $\sigma$ axis.

% Dispersion
%%%%%%%%%%%%%%%%%%%%%%%%%%%%%%%%%%%%%%%%%%%%%%%%%%%%%%%%%%%%%%%%%%%%%%%%%%%%%%%%%%%%%%%%%%%%
\subsection{Dynamical properties}
\label{subsec:dyn_properties}

\begin{figure*}[t!]
	\centerline{\includegraphics[width=18cm]{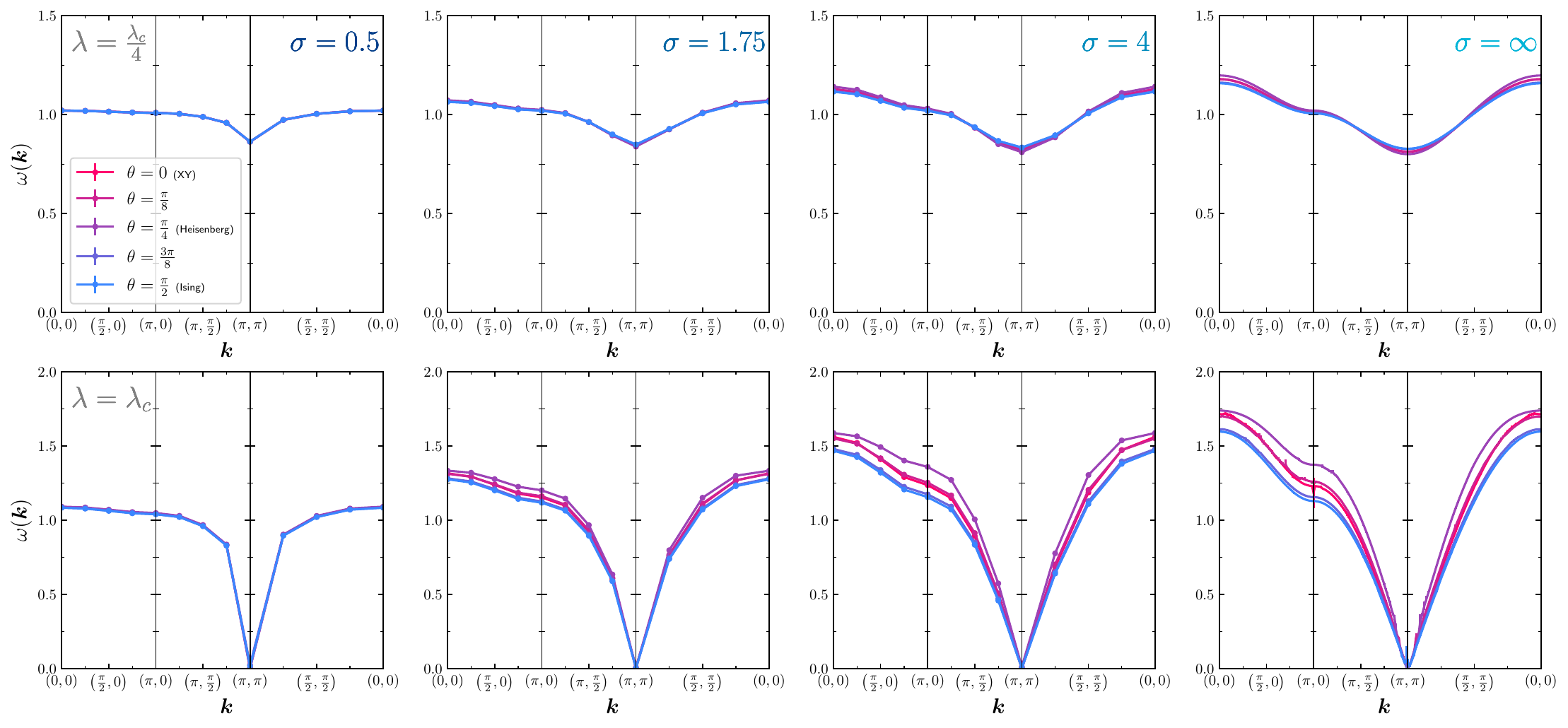}}
	\caption{Dispersion in  the rung-singlet phase evaluated at $\lambda=\lambda_{\rm c}/4$ (upper panels) and at the critical point $\lambda=\lambda_{\rm c}$ (lower panels) for different decay exponents $\sigma\in\{0.5,1.75,4,\infty\}$ and anisotropy parameters $\theta\in\{0,\pi/8,\pi/4,3\pi/8,\pi/2\}$. 
Only the low-energy bands relevant for the gap closing are shown. 
For $0\le\theta\le\pi/4$ the dispersions associated with the triplon excitations $\ket{t^\pm}$ are plotted and for $\pi/4\le\theta\le\pi/2$ the $\ket{t^0}$ dispersion is displayed. 
(\textit{Upper panels}): At $\lambda=\lambda_{\rm c}/4$ the dispersion is gapped and qualitatively shows the same behavior for different $\theta$. 
The behavior about the critical momentum $\bm{k}_{\rm c}=(\pi,\pi)$ changes from quadratic to sublinear upon decreasing $\sigma$. 
(\textit{Lower panels}): At the critical point $\lambda_{\rm c}$ the dispersion becomes gapless for $\bm{k}_{\rm c}=(\pi,\pi)$ and is linear around $\bm{k}_{\rm c}$ for $\sigma\in\{4,\infty\}$. 
At $\sigma=0.5$ within the LRMF regime the dispersions lie perfectly on top of each other and the behavior around $\bm{k}_{\rm c}$ becomes apparently steeper.}
	\label{fig:fig4}
\end{figure*}

After the analysis of the 1qp gap and spectral weight we have a closer look at the 1qp dispersion $\omega^{\text{1qp}}_{\alpha}(\bm{k})$.
Again, we sample the quantity in the three regimes (LRMF, non-trivial, NN) at  $\sigma\in\{0.5,1.75,4,\infty\}$. 
To this end we have to perform MC runs for each momentum $\bm{k}$ of the dispersion. 
For NN interactions, we can apply conventional graph embedding such that the dispersion is available as a high-order series and function in $\lambda$ as well as $\bm{k}$. 
Since triplon excitations carry three different flavors $\alpha\in\{-,0,+\}$, there are three 1qp bands. 
In the isotropic Heisenberg limit $\theta=\pi/4$ these bands are always degenerate. 
Away from the isotropic point the two bands with $\alpha\in\{-,+\}$ must always be degenerate due to the remaining $U(1)$ symmetry (cf.~Eq.~\eqref{eq:anisotropic_int}). 
For $0\le \theta < \pi/4$, these two bands contain the one-triplon gap while the gap is part of the band with $\alpha=0$ for $\pi/4 < \theta \le \pi/2$. 
In App.~\ref{app:bands}, we exemplarily show all bands at $\sigma=\infty$ for all $\theta$.

In the following, we restrict the discussion to the relevant low-energy bands. 
We examine the 1qp dispersion at two distinct values of $\lambda$. 
First, at $\lambda=\lambda_{\rm c}/4$ to gain an understanding deep in the rung-singlet phase and second, the more interesting case, at $\lambda=\lambda_{\rm c}$ to determine the critical behavior. 
In Fig.~\ref{fig:fig4} we present the low-energy 1qp bands for these two perturbation parameter values.

In the upper panels of Fig.~\ref{fig:fig4} we show the 1qp bands for different $\theta$ and $\sigma$ at $\lambda=\lambda_{\rm c}/4$. 
We obtain the dispersion by simply evaluating the gap series at the perturbation parameter value because it is well converged in this regime. 
We observe that the gap remains quite large $\Delta\gtrsim 0.8$ for all $\sigma$. 
Around the critical momentum $\bm{k}_{\rm c}=(\pi,\pi)$ the dispersion is quadratic in the NN regime and it looks like it becomes increasingly steeper for smaller $\sigma$ values away from the NN regime. 
Further, the qualitative behavior of the dispersion is the same for all $\theta$. 
While for larger values of $\sigma$ we can see a small quantitative difference for different $\theta$, at $\sigma=0.5$ the bands all lie almost exactly on top of each other. 
Note, the dispersion for the nearest-neighbor Heisenberg bilayer has been determined by series expansions before \cite{Gelfand1996b, Zheng1997, Collins2008}.

In the lower panels of Fig.~\ref{fig:fig4} we show the 1qp low-energy bands at the critical point $\lambda=\lambda_{\rm c}$. 
Here, we cannot simply evaluate the series but must extrapolate it using DlogPadé approximants to obtain reliable values so far away from the unperturbed limit. 
The difference between the dispersions for different $\theta$ is much more pronounced away from the critical momentum $\bm{k}_{\rm c}$ than before but qualitatively they are still similar. 
For $\sigma=\infty$ some uncertainties from averaging over several DlogPadé approximants are visible as it was not possible to clearly distinguish defective DlogPadés from non-defective ones. 
This problem was more common around the critical momentum, however. In general, the extrapolation works very reliably. 
We again observe at $\sigma=0.5$ in the LRMF regime that the dispersions for different $\theta$ values lie on top of each other. 

Most importantly, we see that the gap closes at \mbox{$\bm{k}_{\rm c}=(\pi,\pi)$} for all $\sigma$ and $\theta$. 
In vicinity of $\bm{k}_{\rm c}$, the dispersion goes linearly to zero in the NN regime. 
For $\sigma=0.5$, the dispersion closes sublinearly around $\bm{k}_{\rm c}$ when looking at the dispersions but it is not clearly visible due to the coarse sampling in the plot. 
In fact, knowing the behavior of the dispersion around the critical gap momentum at the critical point $\lambda_{\rm c}$, we can determine the dynamical critical exponent $z$ by the relation
\begin{equation}
	\left.\omega_{\alpha}^{\text{1qp}}\right\vert_{\lambda=\lambda_{\rm c}} \sim |\bm{k}-\bm{k}_{\rm c}|^{z}.
\end{equation}
If the critical point is not known exactly, extracting $z$ from this relation can be extremely challenging \cite{Kamfor2013PhD}. 
We show that we experience similar difficulties in the NN regime but we can accurately determine $z$ for small enough $\sigma$. 
To obtain the dynamical exponent $z$ we calculate the perturbative series for six different momenta along the path \mbox{$(\pi,\pi)\rightarrow(0,0)$} close to $\bm{k}_{\rm c}=(\pi,\pi)$. 
We extrapolate the series with DlogPadés to determine the dispersion at the critical point $\lambda_{\rm c}$.
Using a linear fit function on a double-log scale we can then extract $z$. 
In App.~\ref{app:fitting_z} we show two representative plots with the linear fit function on double-log scale for $\sigma=0.5$ and $\sigma=4$. 

\begin{figure}[t!]
	\centering
	\includegraphics[width=1.\columnwidth]{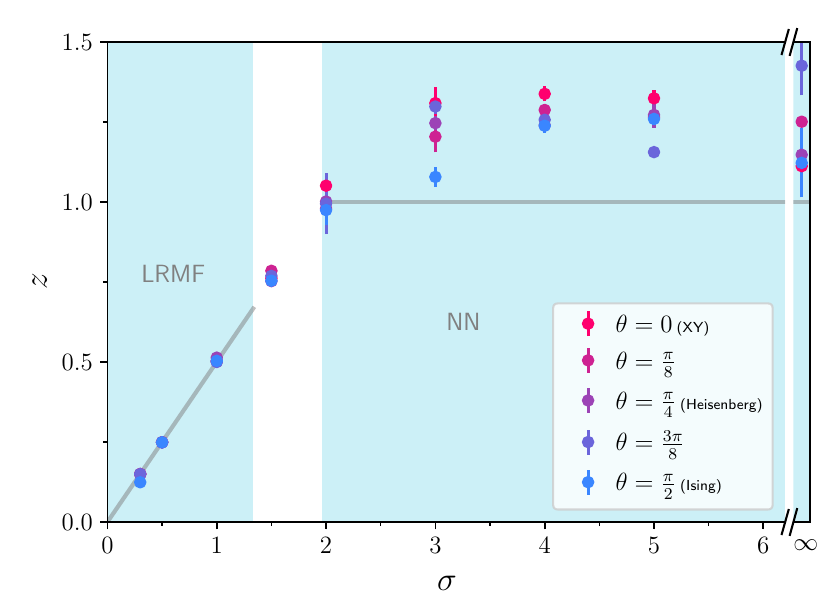}
	\caption{Dynamical critical exponent $z$ as a function of the long-range decay exponent $\sigma$ for different values of the anisotropy parameter $\theta\in\{0,\pi/8,\pi/4,3\pi/8,\pi/2\}$. 
The NN and LRMF regimes are visualized by the blue background and the expected values of $z$ by gray lines. 
In the NN regime and for $\sigma\gtrsim2$ the expected value of $z=1$ is overestimated and the data is scattered for different $\theta$. 
In contrast, for $\sigma\le 2$ the linear behavior of $z$ is captured very well with little deviations.}
	\label{fig:fig5}
\end{figure}

The results for the $z$ exponent as a function of the long-range decay exponent $\sigma$ are depicted in Fig.~\ref{fig:fig5}.
Again, from the long-range $O(n)$ quantum rotor model we exactly know the values of $z$ in the LRMF and NN regime. 
Between the two regimes we expect almost linear behavior of $z$ continuing the straight line of $z$ from the LRMF regime \cite{Defenu2017}. 
For $\sigma\ge 3$ we systematically overestimate the expected constant value of $z=1$ by $\lesssim30\%$ for all $\theta$. 
For a given $\theta$ the values fluctuate considerably with $\sigma$. 
Interestingly, at the interface to the non-trivial regime at $\sigma\sim2$ the estimates for $z$ are all very close to $z=1$. 
From here on into the non-trivial and LRMF regime we capture the expected linear behavior of $z$ with high accuracy. 
The data points are lying almost exactly on the expected line. 
It is not clear to us why, on the one hand $z$ only qualitatively coincides with the expected value of $z=1$ with a systematic offset for $\sigma\ge 3$, and on the other hand is almost perfectly in line with the expectations for $\sigma\le 2$.

Another interesting aspect of determining $z$ is that we can infer the dynamical properties of the 1qp dispersion about the gapless point in the ordered phase from high-order series expansions coming from the disordered limit if the phase on the other side is indeed gapless. 
This is because the dispersion deep in the ordered phase must be adiabatically connected to the critical point and its low-energy properties must not be altered within the same phase. 
We calculate the 1qp dispersion in the ordered phase using a linear spin-wave approximation (SWA) and determine the low-energy behavior $\omega\sim|\bm{k}-\bm{k}_c|^s$ characterized by the exponent $s$ of the gapless Goldstone modes in case of continuous symmetry breaking for $0\le\theta \le \pi/4$. 
See App.~\ref{app:swa} for details on the spin-wave (SW) calculations and App.~\ref{app:goldstone_modes} for the discussion of the results.
Our results corroborate similar results \cite{Diessel2023, Song2023b} on the square lattice Heisenberg model confirming the existence of a sublinear regime $s<1$ for $\sigma<2$ referred to as `anomalous' Goldstone regime. 
In this regime we find that the results for the critical exponent $z$ as well as the dispersion exponent $s$ coincide very well, strengthening the above assessment about the connection of $z$ and $s$ (see Fig.~\ref{fig:fig9} in App.~\ref{app:goldstone_modes}).
It was conjectured that upon entering the superextensive regime $\sigma\le0$, excitations become gapped via a generalized Higgs mechanism \cite{Diessel2023}.
The QMC results in Ref.~\cite{Song2023b} for the  square lattice Heisenberg model indicated that for staggered antiferromagnetic long-range interactions this exotic Higgs regime may already be present above this boundary at $\sigma=0.2$. 
While the convergence of the SW calculations seems to slow down significantly already for $\sigma\le 0.5$, within the pCUT+MC approach the MC sums for points away from $\bm{k}_{\rm c}=(\pi,\pi)$ converge considerably worse only in the regime $\sigma\le0.2$. 
Thus, at least up to $\sigma=0.3$ we do not observe any discrepancies from the expected anomalous gapless behavior.

% Conclusion
%%%%%%%%%%%%%%%%%%%%%%%%%%%%%%%%%%%%%%%%%%%%%%%%%%%%%%%%%%%%%%%%%%%%%%%%%%%%%%%%%%%%%%%%%%%% 
\section{Conclusions}  
\label{sec:conclusions}

In this work, we investigated the antiferromagnetic XXZ square lattice bilayer model with algebraically decaying long-range interactions utilizing the pCUT+MC approach which combines series expansions with classical Monte Carlo integration to tackle systems with long-range interactions. 
Using the isolated rung dimers as the unperturbed limit, we obtained high-order series expansions for the 1qp gap and the 1qp spectral weight in the rung-singlet phase. 
This allowed us through appropriate extrapolation techniques to locate the quantum-critical breakdown of the rung-singlet phase and determine the associated critical exponents. 
We found compelling evidence for three distinct scenarios of $SU(2)$, $U(1)$, and $\mathbb{Z}_2$ symmetry breaking depending on the XXZ anisotropy $\theta$ and three critical regions (LRMF, non-trivial, and NN regime) as a function of the decay exponent $\sigma$. 

We also determined the 1qp dispersion quantitatively up to the quantum phase transition for different $\theta$ and $\sigma$. 
Analyzing the low-energy behavior at the critical point close to the critical momentum, we found that the dynamical exponent $z$ is constant in the NN regime and changes linearly to zero for $\sigma\lesssim 2$ which is not only in agreement with  field-theoretical expectations from the $O(n)$ quantum rotor models \cite{Fisher1972, Sak1973, Sak1977, Dutta2001, Defenu2015, Defenu2017, Defenu2020} but in case of continuous symmetry breaking also with the anomalous behavior of Goldstone modes for $\sigma\lesssim 2$ \cite{Diessel2023}.

We expect that this work will prompt numerous further studies on similar systems since XXZ models (including XY, Heisenberg and Ising interactions) provide a rich and widely unexplored playground for the study of quantum systems with long-range interactions. 
For example, a deeper understanding of the influence of long-range interactions on the amplitude (Higgs) mode in the ordered phase of Heisenberg models \cite{Powalski2015, Powalski2018, Lohoefer2015} is highly desirable, especially in the context of altered Goldstone modes and a gap opening via a generalized Higgs mechanism for strong long-range interactions \cite{Diessel2023}. 
The interplay of frustration and long-range Heisenberg interactions is another poorly understood domain of research which is believed to potentially give rise to exotic quantum spin liquid ground states in the superextensive regime as conjectured in Ref.~\cite{Chiocchetta2021}. 
Therefore, modifying and generalizing our series expansion approach to handle strong long-range interactions and enable the study of bound states as well as arbitrary lattice geometries is a promising route for the future. 

\section*{Acknowledgments}
P.A. thanks Max Hörmann for fruitful discussions. P.A., A.D., and K.P.S. gratefully acknowledge the support by the Deutsche Forschungsgemeinschaft (DFG, German Research Foundation) -- Project-ID 429529648—TRR 306 \mbox{QuCoLiMa} (``Quantum Cooperativity of Light and Matter'') and the Munich Quantum Valley, which is supported by the Bavarian state government with funds from the Hightech Agenda Bayern Plus. The authors gratefully acknowledge the scientific support and HPC resources provided by the Erlangen National High Performance Computing Center (NHR@FAU) of the Friedrich-Alexander-Universität Erlangen-Nürnberg (FAU) under the NHR project b177dc (``SELRIQS''). NHR funding is provided by federal and Bavarian state authorities. NHR@FAU hardware is partially funded by the German Research Foundation (DFG) – 440719683.

\section*{Data availability}
The raw data is available on Zenodo at: \mbox{\url{https://doi.org/10.5281/zenodo.14213076}}.
The code used to generate the numerical results presented in this paper can be made available by Patrick Adelhardt (patrick.adelhardt@fau.de) upon reasonable request.

\section*{Author contributions}
P.A.: Conceptualization, Data curation (pCUT+MC), Formal analysis (all), Investigation (pCUT+MC), Methodology (pCUT+MC), Writing -- original draft, Writing -- review \& editing. A.D.: Data curation (SWA), Formal analysis (SWA), Investigation (SWA), Writing -- original draft (App.~\ref{app:swa}, \ref{app:goldstone_modes}), Writing -- review \& editing. K.P.S.:  Conceptualization, Supervision, Writing -- review \& editing. Following the taxonomy \href{https://credit.niso.org/}{CRediT} to categorize the contributions of authors.

\appendix

\section{Quasiparticle processes of the Hamiltonian and observables}
\label{app:processes}

Every perturbation problem can essentially be brought into the form  $\mathcal{H} = \mathcal{H}_0 + \lambda\mathcal{V}$, where $\mathcal{H}_0$ is the unperturbed part, $\lambda$ the perturbation parameter, and $\mathcal{V}$ the perturbation. 
We reformulate the Hamiltonian of the XXZ bilayer model with long-range interactions given by Eqs.~\eqref{eq:ham_bilayer} and \eqref{eq:anisotropic_int} in terms of rung singlet and triplet excitations. 
Then the perturbation has the structure 
\begin{equation}
	\mathcal{V} = T_{-2} + T_0 + T_{+2},
\end{equation}
where the $T_n$-operators contain all processes changing the quasiparticle number (triplet number) by $n$. 
We can further decompose the perturbation by writing
\begin{equation}
	T_{n} = \sum_{l} \tau_{l,n}\,,
\end{equation}
decomposing $T_n$ into local operators $\tau_{l,n}$, where $l$ can be the index of any bond on the lattice. 
Thus, the local operator acts on a bond and therefore on a pair of rung dimers each being in one of the singlet or triplet states. 
In Tab.~\ref{tab:ham_processes} all possible processes are listed. 
\begin{table}
	\caption{Quasiparticle processes for the Hamiltonian in Eq.~\eqref{eq:ham_bilayer} are listed for the local operators $\tau_{0}$ and $\tau_{\pm2}$. Processes depend on the XXZ anisotropy parameter $\theta$ where processes in the left column have the prefactor $\sin\theta$ and the ones in the right column have $\cos\theta$ as a prefactor. The arrows $\leftrightarrow$ signify that processes can occur in both directions and $\rightarrow$ that processes only have one direction from left to right. 
The operators in the table are multiplied by a factor of $2$ for the sake of compact notation.}
	\begin{tabular}{cc}
    \hline\hline
    \multicolumn{2}{c}{$2\tau_0$} \\
    \hline
    $\sin\theta$ &
    $\cos\theta$ \\
    \begin{tabular}[t]{@{}l@{}}
        $\ket{t^{0},s} \leftrightarrow \ket{s, t^{0}}$ \\
        $\ket{t^{\pm},t^{\pm}} \rightarrow \ket{t^{\pm}, t^{\pm}}$ \\
        $\ket{t^{\pm},t^{\mp}} \rightarrow -\ket{t^{\pm}, t^{\mp}}$
    \end{tabular} &
    \begin{tabular}[t]{@{}l@{}}
        $\ket{t^{\pm},s} \leftrightarrow \ket{s, t^{\pm}}$ \\
        $\ket{t^{0},t^{\pm}} \leftrightarrow \ket{t^{\pm}, t^{0}}$ \\
        $\ket{t^{\pm},t^{\mp}} \leftrightarrow \ket{t^0,t^0}$
    \end{tabular} \\
    \multicolumn{2}{c}{$2\tau_{+2}$} \\
    \hline
    $\sin\theta$ &
    $\cos\theta$ \\
    $\ket{s,s} \rightarrow \ket{t^0,t^0}$ &
    $\ket{s,s} \rightarrow -\ket{t^+,t^-} - \ket{t^-,t^+}$ \\
    \hline\hline
\end{tabular}
	\label{tab:ham_processes}
\end{table}

The operator $\tau_0$ contains only quasiparticle-conserving processes and the operator $\tau_{+2}$ contains all processes where two triplet quasiparticles are created. 
The operator $\tau_{-2}$ is related to $\tau_{+2}$ via $\tau_{+2}^{\dagger} = \tau_{-2}^{\phantom{\dagger}}$. 

To determine the 1qp spectral weight (see Eq.~\eqref{eq:1qp_sw}) of the antisymmetric spin observables \eqref{eq:spin_obs}, we also need to determine their action in terms of singlet and triplet excitations. 
We consider only antisymmetric combinations of spin operators as they couple to odd quasiparticle numbers and we here want to calculate the 1qp spectral weight. 
This is straightforward and we list the action of $\mathcal{O}_i^{z}$ and $\mathcal{O}_i^{x}$ in Tab.~\ref{tab:obs_processes}.
\begin{table}
	\caption{Quasiparticle processes for the 1qp spectral weight of the spin observables \eqref{eq:spin_obs} are listed. 
The arrows $\leftrightarrow$ signify that processes can occur in both directions and $\rightarrow$ that processes only have one direction from left to right. 
The observables in the table are multiplied by a factor of $2$ for the sake of compact notation.}
	\begin{tabular}{c}
		\hline\hline
		$2\mathcal{O}^z=\left(S_{1}^z-S_{2}^z\right)$ \\
		\hline
		$\begin{gathered}
			\begin{aligned}
				&\ket{s}\leftrightarrow \ket{t^0}\\
				&\ket{t^{\pm}} \rightarrow 0 \\
			\end{aligned}
		\end{gathered}$\\
		$2\mathcal{O}^x=\sqrt{2}\left(S_{1}^x-S_{2}^x\right)$ \\
		\hline
		$\begin{gathered}
			\begin{aligned}
				&\ket{s} \leftrightarrow \mp \ket{t^{\pm}} \\
				&\ket{t^0}\rightarrow 0\\
			\end{aligned}
		\end{gathered}$\\
		\hline\hline
	\end{tabular}
	\label{tab:obs_processes}
\end{table}
Note, the spin observables
\begin{align}
	\mathcal{O}^{x}_i &=  \frac{1}{\sqrt{2}}\left(S^x_{i,1}-S^x_{i,2}\right)\, ,\\
	\mathcal{O}^{z}_i &=  \frac{1}{2}\left(S_{i,1}^z-S_{i,2}^z\right)
\end{align}
differ by a factor of $1/\sqrt{2}$ as we want to account for the fact that the observable $\mathcal{O}^x$ couples a singlet $\ket{s}$ to two triplets $\ket{t^{\pm}}$ mixing the quasiparticle contributions, while $\mathcal{O}^z$ couples the singlet to one triplet $\ket{t^0}$. 

\section{One-triplon bands}
\label{app:bands}

\begin{figure}[t!]
	\centering
	\includegraphics[width=1.\columnwidth]{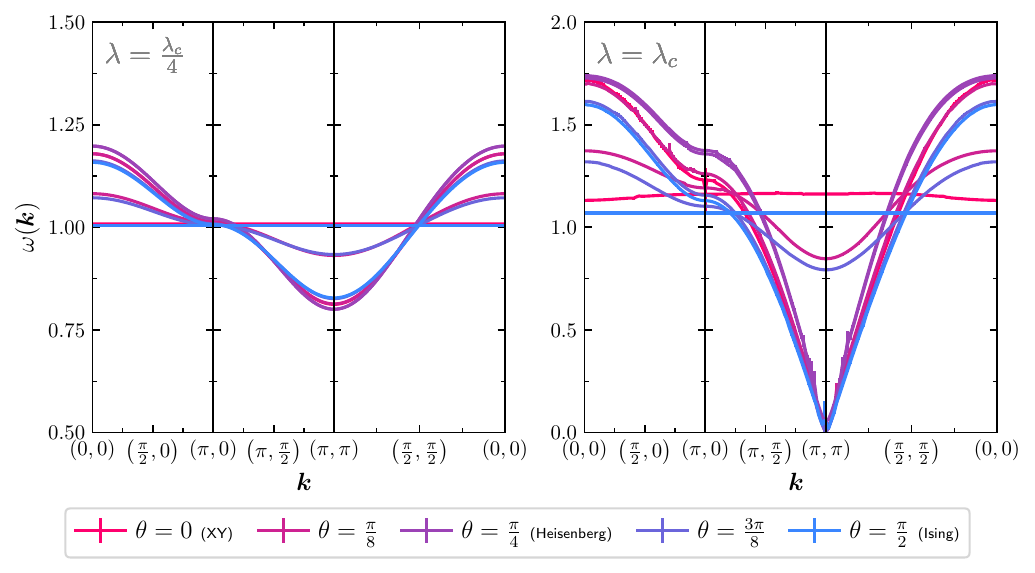}
	\caption{1qp bands for various $\theta\in\{0,\pi/8,\pi/4, 3\pi/8, \pi/2\}$ evaluated at $\lambda=\lambda_{\rm c}/4$ (left panel) and at $\lambda=\lambda_{\rm c}$ (right panel). 
The critical bands associated with the relevant triplet excitations which condense at the critical point as well as the non-critical bands are shown.}
	\label{fig:fig6}
\end{figure}

In Fig.~\ref{fig:fig4} in the main text we showed the 1qp dispersion for $\sigma\in\{0.5,1.75,4,\infty\}$ evaluated at $\lambda=\lambda_{\rm c}/4$ (upper panels) and at $\lambda=\lambda_{\rm c}$ (lower panels). 
However, the plot only includes the low-energy 1qp bands which close at the quantum-critical point $\lambda_{\rm c}$ while the non-critical bands are not included. 
In the region $0\le\theta<\pi/4$ the triplet excitations $\ket{t^{\pm}}$ are the relevant quasiparticles which condense at the quantum-critical point, such that the gap of the two degenerate bands associated with these excitations closes. 
In the other region $\pi/4<\theta\le\pi/2$ the triplet excitation $\ket{t^0}$ is the critical one and its associated band closes at the critical point. 
At the $SU(2)$-symmetric Heisenberg point at $\theta=\pi/4$, all 1qp bands are degenerate and close at the critical point.
In Fig.~\ref{fig:fig6} we show all bands for $\sigma=\infty$ for different anisotropies $\theta\in\{0,\pi/8,\pi/4, 3\pi/8, \pi/2\}$ evaluated at $\lambda=\lambda_{\rm c}/4$ (left panel) and at $\lambda=\lambda_{\rm c}$ (right panel).
In the limiting cases $\theta=0$ ($\theta=\pi/2$) we can see two (one) dispersive bands and one (two) flat bands in the rung-singlet phase at $\lambda=\lambda_{\rm c}/4$. 
Upon tuning from the limiting cases towards the Heisenberg point, processes mixing triplet flavors become increasingly relevant and the non-critical bands that were previously flat become dispersive, as can be seen e.\,g. for the non-critical bands for $\theta=\pi/8$ and $\theta=3\pi/8$. 
At $\lambda=\lambda_{\rm c}$ the bands associated to the relevant triplet excitation become zero at $\bm{k}=(\pi,\pi)$ as expected from the antiferromagnetically ordered phases for $\lambda>\lambda_{\rm c}$. 
Interestingly, the non-critical band in the limiting case $\theta=0$ becomes slightly dispersive. 
Looking at Tab.~\ref{tab:ham_processes} this becomes clear as even though all sine terms become zero there are still processes converting $\ket{t^0}$ to $\ket{t^{\pm}}$ triplets, which allows the $\ket{t^0}$ triplets to still be mobile. 
In the other limiting case $\theta=\pi/2$ there are no such processes converting the triplet flavors and hence the non-critical band remains flat.

We also tried to calculate the non-critical 1qp bands using the pCUT+MC approach for $\sigma\in\{0.5,1.75,4\}$, however, we found that the series coefficients do not converge. 
It is not clear to us whether this is due to insufficiencies of the MC moves or if there are any fundamental physical reasons that are not obvious. Certainly, further investigations are needed in the future for clarification.

\section{Convergence of DlogPadé approximants}
\label{app:convergence}

We discuss the convergence behavior of (biased) DlogPadé approximants for the 1qp excitation gap $\Delta$ (spectral weight $\mathcal{S}^{\rm 1qp}_{\alpha}$) as a function of the long-range decay exponent $\sigma$ and the anisotropy $\theta$. We already introduced our extrapolation scheme in Sec.~\ref{subsec:extrapolation} with the definition of (biased) Padé approximants $P[L,M]_{\mathcal{D}}$ ($P[L,M]_{\vartheta^{\star}}$) of the logarithmic derivative $\mathcal{D}$ given by Eq.~\eqref{eq:dlogpade} (Eq.~\eqref{eq:biaseddlogpade}). The associated DlogPadé approximants can be obtained upon integration. By analyzing the poles of the approximant $P[L,M]_{\mathcal{D}}$ we can identify the physical pole which determines the critical point $\lambda_c$ and from its residuum we can determine the associated critical exponent. It is of great importance to exclude spurious extrapolants with non-physical poles in the complex plane close to the real axis for $\lambda<\lambda_c$. This also holds for the biased approximant $P[L,M]_{\vartheta^{\star}}$. All non-spurious DlogPad\'{e} approximants are grouped into families with $L-M = {\rm const.}$ and $-3\le L-M \le 3$. They are expected to converge to the true value for the approximants' order $\frak{o}\rightarrow\infty$. To get a good estimate of $\lambda_c$ and $z\nu$, we simply average over the highest order approximants from each family. The error bars in the figures of the main text are therefore given by the standard deviation from averaging over approximants.
\begin{figure}[t!]
	\centering
	\includegraphics[width=1.\columnwidth]{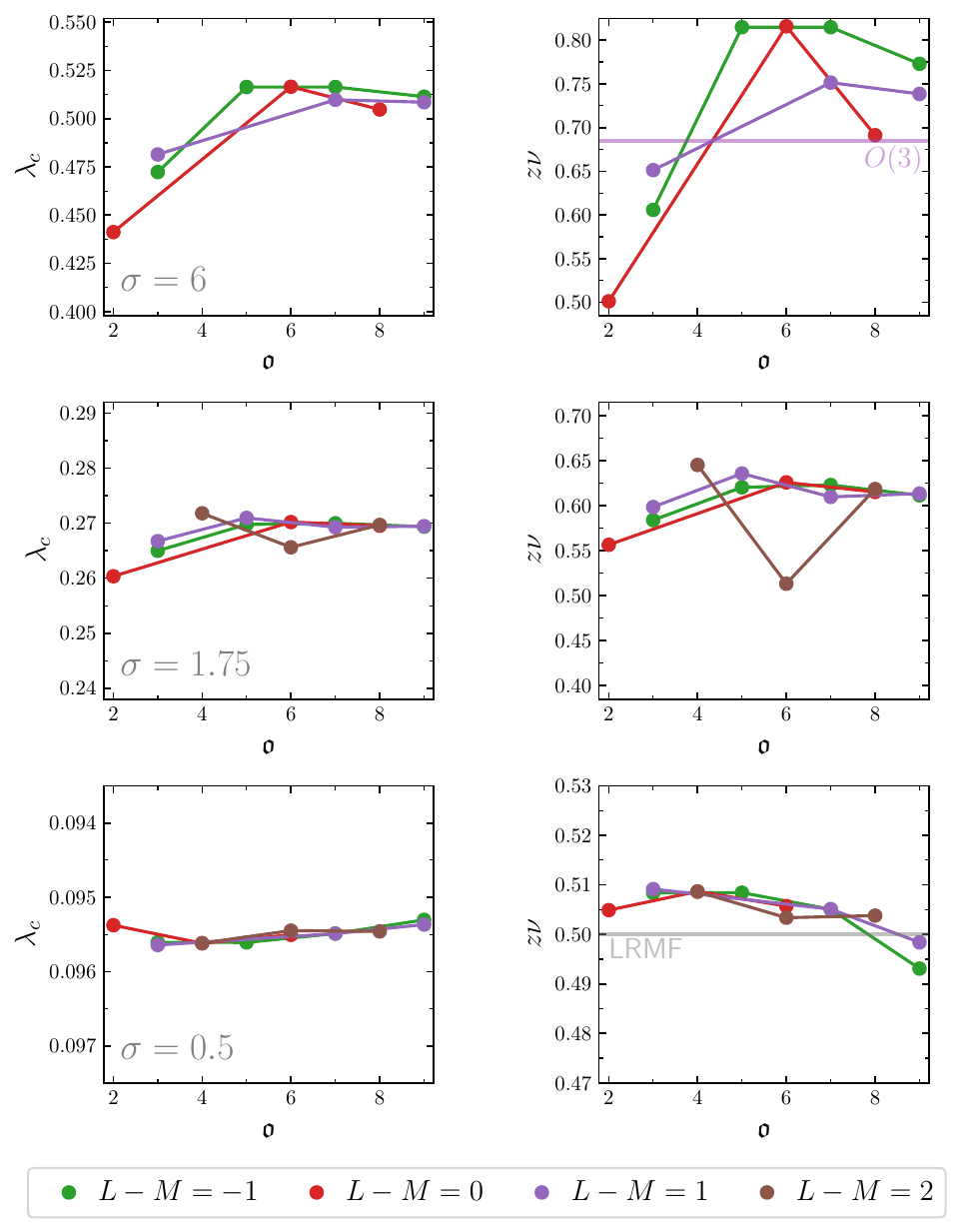}
	\caption{Estimates for the critical point $\lambda_c$ (left column) and critical exponent $z\nu$ (right column) from DlogPadé approximants of the 1qp gap series $\Delta$ for $\sigma\in\{0.5, 1.75, 6\}$ and Heisenberg interactions ($\theta=\pi/4$). The approximants are grouped into families with $L-M = {\rm const.}$, $-3\le L-M \le 3$ and minimal family length 3 and the non-spurious ones are plotted as a function of the approximants' order $\frak{o}=L+M<\frak{o}_{\rm max}$, where $L$ is the order of the numerator, $M$ the order of the denominator and $\frak{o}_{\rm max}=10$ the order of the series. Horizontal lines show literature values of critical exponents for reference.}
	\label{fig:fig7}
\end{figure}

In Fig.~\ref{fig:fig7} we show the critical point $\lambda_c$ and the critical exponent $z\nu$ from DlogPadé approximants of the 1qp gap series $\Delta$ for $\sigma\in\{0.5, 1.75, 6\}$ as a function of the approximants' order $\frak{o}=L+M<\frak{o}_{\rm max}=10$, where $L$ is the order of the numerator, $M$ the order of the denominator, and $\frak{o}_{\rm max}$ the order of the series. For the unbiased approximants we chose a minimal family length of 3. We only show here the results for the Heisenberg case $\theta=\pi/4$ as the approximants exhibit the worst convergence behavior for this interaction type. Note, the obtained estimates for the critical point $\lambda_c$ are usually one order of magnitude more precise than for the associated critical exponents as can be seen in the figure. For $\sigma=6$ in the nearest-neighbor regime, the variance between different families is the largest and the deviation from the literature value of the $O(3)$ critical exponent is considerable but in line with previous series expansion results for nearest-neighbor interactions \cite{Coester2011Diploma}. In contrast, for $\sigma=0.5$ in the long-range mean-field regime, the approximant families are close to the literature value of the critical exponent and the variance is much smaller (the plot range is one order of magnitude smaller). Indeed, we observe similar convergence behavior for any anisotropy $\theta$. Overall, our pCUT+MC approach yields better estimates of critical properties the stronger the long-range interactions.

\begin{figure}[t!]
	\centering
	\includegraphics[width=1.\columnwidth]{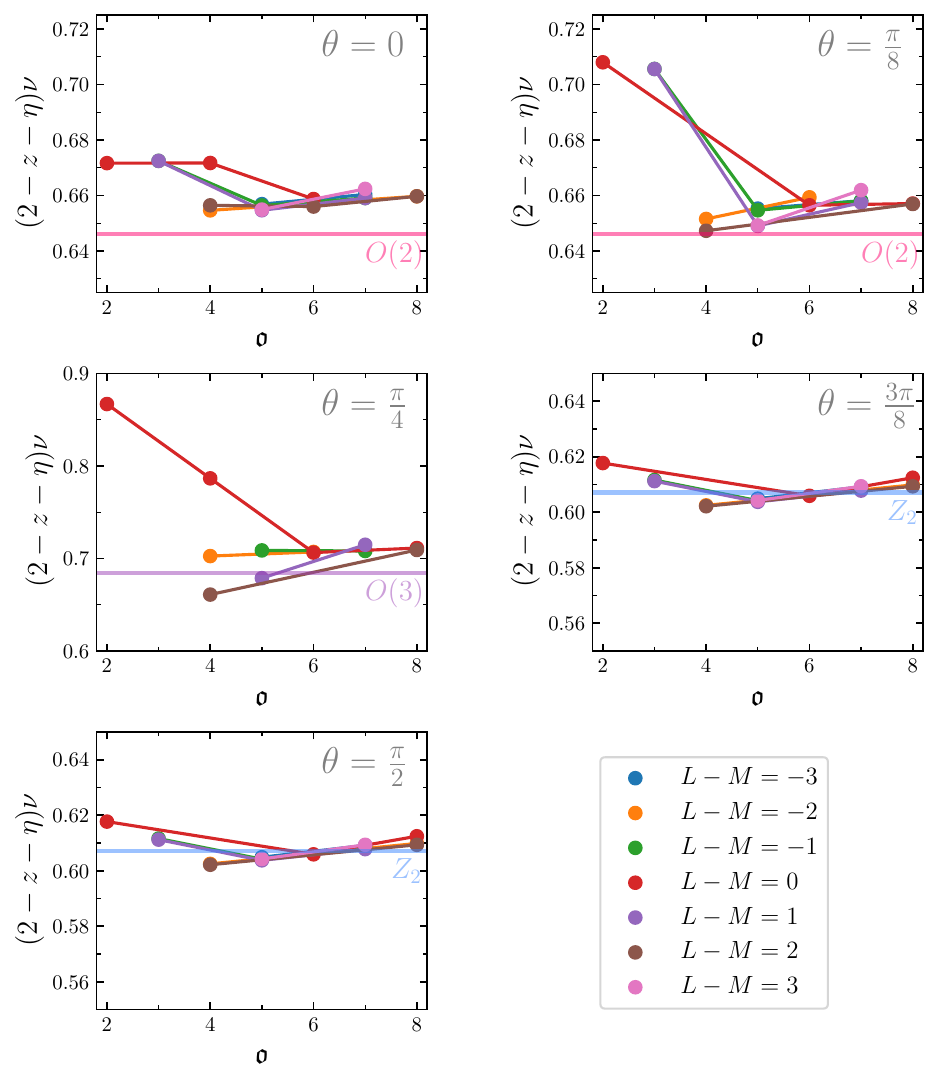}
	\caption{Estimates for the  critical exponent $(2-z-\eta)\nu$ from biased DlogPadé approximants of the 1qp spectral weight series $\mathcal{S}^{\rm 1qp}_{\alpha}$ for $\theta\in\{0, \pi/8, \pi/4, 3\pi/8, \pi/2\}$ and  $\sigma=6$ in the nearest-neighbor regime. The estimate for $\lambda_c$ from the 1qp gap closing was used to bias the approximants. The approximants are grouped into families with $L-M = {\rm const.}$, $-3\le L-M \le 3$ and minimal family length 2 and plotted as a function of the approximants' order $\frak{o}=L+M<\frak{o}_{\rm max}$, where $L$ is the order of the numerator, $M$ the order of the denominator, and $\frak{o}_{\rm max}=9$ the order of the series. Horizontal lines show literature values of critical exponents for reference.}
	\label{fig:fig8}
\end{figure}
In Fig.~\ref{fig:fig8} we plot the critical exponent $(2-z-\eta)\nu$ from biased DlogPadé approximants of the 1qp spectral weight $\mathcal{S}^{\rm 1qp}_{\alpha}$ for $\theta\in\{0, \pi/8, \pi/4, 3\pi/8, \pi/2 \}$ at a fixed decay exponent $\sigma=6$ as a function of the approximants' order $\frak{o}=L+M<\frak{o}_{\rm max}=9$ with a minimal family length of 2. We use the critical point $\lambda_c$ obtained from the closing of the 1qp gap since determining $\lambda_c$ from the closing of a quantity is more accurate than from the divergence of a quantity (gap closing versus divergence of the spectral weight in the critical regime). Therefore, biased DlogPadés yield significantly better results for the critical exponent $(2-z-\eta)\nu$. We observe that the convergence of the data is the best in the Ising regime ($\theta>\pi/4$) while the largest deviations occur for Heisenberg interactions ($\theta=\pi/4$) (note the different plot ranges). Also, it is clearly visible that DlogPadé approximants tend to slightly overestimate critical values. 

\vspace{1em}
\section{Determining the dynamical exponent $z$}
\label{app:fitting_z}

\begin{figure}[t!]
	\centering
	\includegraphics[width=1.\columnwidth]{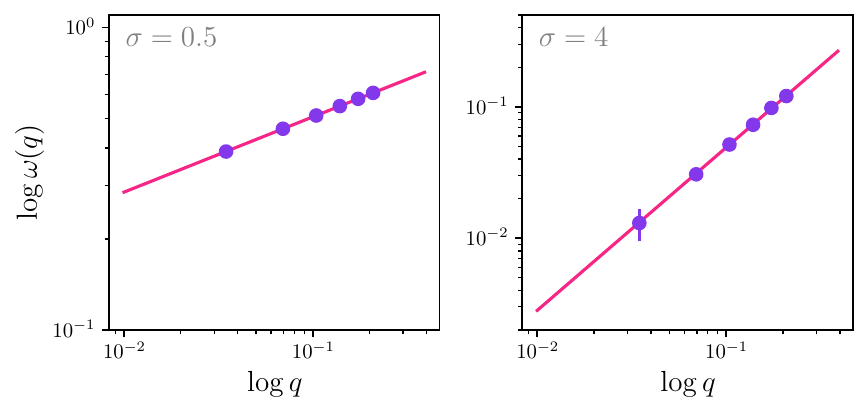}
	\caption{Linear fit on double-log scale to determine the critical dynamical exponent $z$ from its power-law behavior at the critical point $\sim |\bm{q}|^z$. 
For  $\sigma=0.5$ we obtain $z=0.24891(24)$ and for $\sigma=4$ we get $z=1.240(16)$.}
	\label{fig:fig9}
\end{figure}

We want to extract the critical dynamical exponent $z$ from the 1qp dispersion using the fact that the dispersion follows a power-law behavior 
\begin{equation}
	\left.\omega_{\alpha}^{\text{1qp}}\right\vert_{\lambda=\lambda_{\rm c}} \sim |\bm{q}|^{z}.
\end{equation}
at the critical point $\lambda_{\rm c}$,  where $\bm{q} = \bm{k}_{\rm c}-\bm{k}$ is the shifted momentum. 
To determine $z$ we sample the dispersion along the path $(\pi,\pi)\rightarrow (0,0)$ close to $\bm{k}_{\rm c}=(\pi,\pi)$. 
Choosing the stepsize $(\pi/128, \pi/128)$, we sample six data points along the path. 
For every sampled $\bm{k}$ close to $\bm{k}_{\rm c}$ we perform MC runs and use DlogPadé extrapolations to evaluate the dispersion at $\lambda=\lambda_{\rm c}$. 
Note, it is necessary to determine $\lambda_{\rm c}$ beforehand by pinpointing the physical pole of the series at $\bm{k}_{\text{c}}$. 
We can now determine the dynamical exponent $z$ by using the fit function
\begin{equation}
	f(q)= aq^z
\end{equation}
where $a$ and $z$ are free fit parameters and $q=\sqrt{q_x^2+q_y^2}$. 
On double-log scale this becomes a linear fit function where $z$ is the slope. 
In Fig.~\ref{fig:fig9} we show two such fits as examples for isotropic Heisenberg interactions ($\theta=\pi/4$) for $\sigma=0.5$ and $\sigma=4$.
The data points follow the suggested power-law behavior well and we experienced for all fits that the data points do not scatter much except for very few exceptions, where it was difficult to distinguish defective from non-defective DlogPadé approximants. 
Just from the quality of the fits, we are not able to explain why the dynamical exponent coincides with the expected behavior almost perfectly for $\sigma\le 2$ and why there is a systematic offset for large decay exponents (compare results shown in Fig.~\ref{fig:fig5}).

\section{Linear spin-wave calculations for the Heisenberg bilayer}
\label{app:swa}

\begin{figure}[t!]
	\centering
	\includegraphics[width=1.\columnwidth]{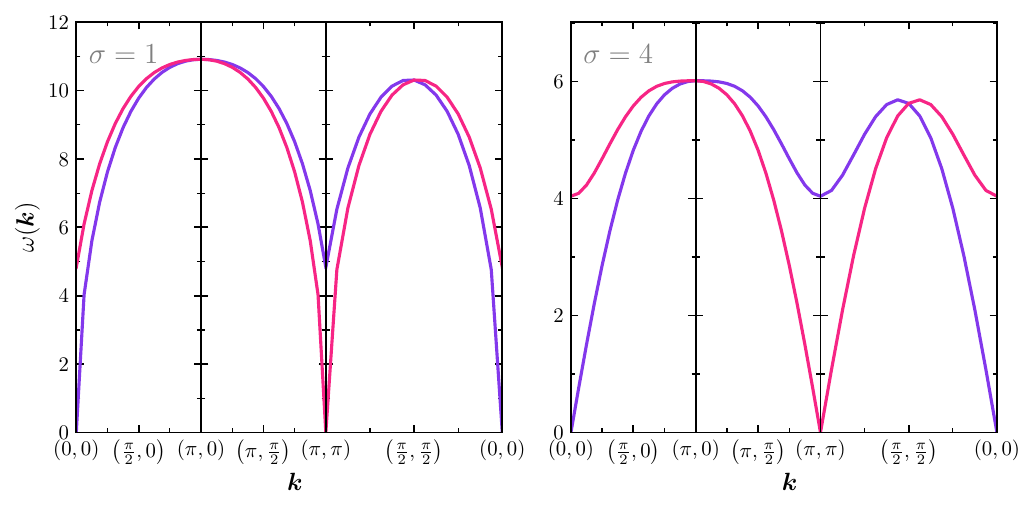}
	\caption{Dispersion $\omega_{+}(\bm{k})$ (purple) and $\omega_{-}(\bm{k})$ (pink) from spin wave calculations for $\sigma=1$ (left) and $\sigma=4$ (right). While for $\sigma=4$ the dispersion is linear around the gapless point $\bm{k}=(\pi,\pi)$, it becomes sublinear when decreasing $\sigma$ to 1. }
	\label{fig:fig10}
\end{figure}

In this section we outline the linear spin wave (SW) calculations for the Heisenberg bilayer model (i.e. $\theta=\pi/4$) with antiferromagnetic interlayer and staggered antiferromagnetic intralayer long-range interactions,
\begin{equation}
	\mathcal{H} = J_{\perp} \sum_{\bm{i}} \vec{S}_{\bm{i}, 1} \vec{S}_{\bm{i}, 2} -  \frac{1}{2}\sum_{n = 1}^{2} \sum_{\bm{i}\neq \bm{j}} J(\bm{j}-\bm{i})\vec{S}_{\bm{i}, n} \vec{S}_{\bm{j}, n}\,
	\label{eq:heisenberg_ham}.
\end{equation}
with $J_{\perp},J>0$ and $J(\bm{\delta})$ as defined in Eq.~\eqref{eq:stag_lr_int} in the main text.
The classical ground state exhibits antiferromagnetic order between nearest neighbors which calls for the definition of two sublattices $A$ and $B$ with spins on different sublattices pointing in opposite directions. 
We use the Holstein-Primakoff transformation to map the spin operators to bosonic operators $a^{(\dagger)}$ and $b^{(\dagger)}$ on sublattices $A$ and $B$, respectively.
In leading order in the bosonic operators we can write the transformation as
\begin{equation}
	\begin{gathered}
		\begin{aligned}
			S_{\bm{i},1}^{z}&= S - a_{\bm{i},1}^{\dagger}a_{\bm{i},1}^{\phantom\dagger} &~~ S_{\bm{i},1}^{-}&\approx \sqrt{2S}a_{\bm{i}, 1}^{\dagger} &~~ S_{\bm{i},1}^{+}&\approx\sqrt{2S}a_{\bm{i},1}^{\phantom\dagger}\,,
		\end{aligned}\\
		\begin{aligned}
			S_{\bm{i},2}^{z}&= b_{\bm{i},2}^{\dagger}b_{\bm{i},2}^{\phantom\dagger} - S &~~ S_{\bm{\bm{i}},2}^{-}&\approx \sqrt{2S}b_{\bm{i}, 2}^{\phantom\dagger} &~~ S_{\bm{\bm{i}},2}^{+}&\approx\sqrt{2S}b_{\bm{i},2}^{\dagger}\,,
		\end{aligned}\\
		\begin{aligned}
			S_{\bm{j},1}^{z}&= b_{\bm{j},1}^{\dagger}b_{\bm{j},1}^{\phantom\dagger} - S &~~ S_{\bm{j},1}^{-}&\approx \sqrt{2S}b_{\bm{j},1}^{\phantom\dagger} &~~ S_{\bm{j},1}^{+}&\approx\sqrt{2S}b_{\bm{j},1}^{\dagger}\,,
		\end{aligned}\\
		\begin{aligned}
			S_{\bm{j},2}^{z}&= S - a_{\bm{j},2}^{\dagger}a_{\bm{j},2}^{\phantom\dagger} &~~ S_{\bm{j},2}^{-}&\approx \sqrt{2S}a_{\bm{j},2}^{\dagger} &~~ S_{\bm{j},2}^{+}&\approx\sqrt{2S}a_{\bm{j},2}^{\phantom\dagger}\,.
		\end{aligned}
	\end{gathered}
\end{equation}
Inserting these expressions into the Hamiltonian \eqref{eq:heisenberg_ham}, neglecting the quartic term of bosonic operators and performing a Fourier transformation, yields (in units of $J_{\perp}$) the linear spin-wave Hamiltonian
\begin{widetext}
	\begin{equation}
		\begin{split}
			\mathcal{H}^{\text{SW}} \approx \text{const.} + S \sum_{\bm{k}} \Big\{ &\sum_{n=1}^2
			\left[\left(1+ \gamma +f(\bm{k})\right)\left(a_{\bm{k}, n}^{\dagger}a_{\bm{k}, n}^{\phantom\dagger} + b_{-\bm{k}, n}^{\dagger}b_{-\bm{k}, n}^{\phantom\dagger} \right) 
			+ g(\bm{k})  \left(a_{\bm{k}, n}^{\dagger}b_{\bm{-k}, n}^{\dagger} + a_{\bm{k}, n}^{\phantom\dagger}b_{-\bm{k}, n}^{\phantom\dagger} \right)\right] \\ &+ a_{\bm{k}, 1}^{\phantom\dagger}b_{-\bm{k}, 2}^{\phantom\dagger} 
			+ b_{\bm{k}, 1}^{\dagger}a_{-\bm{k}, 2}^{\dagger} + b_{\bm{k}, 1}^{\phantom\dagger}a_{-\bm{k}, 2}^{\phantom\dagger}  + b_{\bm{k}, 1}^{\dagger}a_{-\bm{k}, 2}^{\dagger}   \Big\}.
		\end{split} % divided by J_perp, we have J_{||} > 0 so \lambda>0
	\end{equation}
\end{widetext}

where the prefactors $f(\bm{k})$, $g(\bm{k})$ and $\gamma$ are defined as
\begin{align} %antiferro case
			\gamma = \lambda& \sum_{\bm{\delta}\in\text{diff}} \frac{1}{|\bm{\delta}|^{2+\sigma}}, \nonumber\\ 
			f(\bm{k}) = -\lambda& \sum_{\bm{\delta}\in\text{same}} \frac{\mathrm{e}^{i\bm{k}\bm{\delta}} -1}{|\bm{\delta}|^{2+\sigma}},  \label{eq:SW_prefactors_af}\\
			g(\bm{k}) = \lambda& \sum_{\bm{\delta}\in\text{diff}} \frac{\mathrm{e}^{i\bm{k}\bm{\delta}}}{|\bm{\delta}|^{2+\sigma}} . \nonumber	
\end{align}
with $\lambda=J/J_{\perp}>0$ and the staggered interactions are accounted for by the additional sign in $f(\bm{k})$. The sums run over all $\bm{\delta}$ within the same or different sublattices, respectively, as indicated.
Following Refs.~\cite{Xiao2009,Adelhardt2023}, we diagonalize the Hamiltonian which is quadratic in the creation and annihilation operators in momenta using a Bogoliubov-Valatin transformation. This yields the diagonal Hamiltonian
\begin{equation}
	\mathcal{H}^{\text{SW}} = \text{const.} + S \sum_{\bm{k},n}\left(\omega_{+}(\bm{k}) \alpha_{\bm{k},n}^{\dagger}\alpha_{\bm{k},n}^{\phantom\dagger} + \omega_{-}(\bm{k})\beta_{\bm{k},n}^{\dagger}\beta_{\bm{k},n}^{\phantom\dagger}\right)
\end{equation}
in terms of new boson creation and annihilation operators $\alpha_{\bm{k},\nu}^{(\dagger)} $ and $\beta_{\bm{k},\nu}^{(\dagger)}$ with the spin-wave dispersion
\begin{equation} % antiferro bilayer
	\omega_{\pm}(\bm{k}) = \sqrt{\left(1+\gamma + f(\bm{k})\right)^2 - \left(g(\bm{k}) \pm 1\right)^2}.
	\label{eq:SW_disp_af}
\end{equation}
We plot the dispersion in Fig.~\ref{fig:fig10} exemplarily for $\sigma=1$ and $\sigma=4$, using $\delta_{\text{max}}=800$ for the evaluation of the prefactors in Eq.~\eqref{eq:SW_prefactors_af} (for details on the evaluation, see description in App.~\ref{app:goldstone_modes}). The band $\omega_{-}$ is gapless at $\bm{k}_c=(\pi,\pi)$ and the plot suggests a linear dispersion around $\bm{k}_c$ for large $\sigma$ which becomes sublinear upon decreasing $\sigma$.  

\section{Dynamical exponent and Goldstone modes in the Heisenberg bilayer model}
\label{app:goldstone_modes}

Here, we discuss the results obtained from the linear spin-wave (SW) analysis for the (staggered) antiferromagnetic square lattice Heisenberg bilayer model.
In App.~\ref{app:swa} we already determined the SW dispersion~\eqref{eq:SW_disp_af} in the ordered phase. In particular, we are interested in the gapless band $\omega_{-}(\bm{k})$ around the gapless point $\bm{k}_{\rm c}=(\pi,\pi)$. 
Analogously to App.~\ref{app:fitting_z} we fit the SW dispersion with a function $\omega\sim|\bm{q}|^s$ with $\bm{q}=\bm{k}_c - \bm{k}$ on a double-logarithmic scale where the fit function becomes linear and we can get the exponent $s$ from the slope. 
We sample $\omega_{-}$ around $\bm{k}_{\rm c}$ with steps of $0.01$ and use the first five data points around $\bm{k}_{\rm c}$ for the linear fit. To determine the SW dispersion $\omega_{-}(\bm{k})$, first we need to calculate the prefactors $f(\bm{k})$, $g(\bm{k})$ and $\gamma$ (see Eqs.~\eqref{eq:SW_prefactors_af}), which due to the long-range interactions all contain an infinite sum over all lattice vectors $\bm{\delta}$. 
Hence, in order to evaluate $\omega$, we truncate these sums such that all $\bm{\delta} = (\delta_x,\delta_y)$ with $\delta_{x,y} \in [-\delta_\text{max},\delta_\text{max}]$ are included. We find that for $\sigma\gtrsim 1$, the results for $s$ converge well with $\delta_\text{max}$, using $\delta_\text{max}=500$ for $\sigma\in[3,6]$ and $\delta_\text{max}=800$ for $\sigma\in[0,3)$. 
For $\sigma\lesssim 1$, we still observe a significant monotonous decrease in the extracted value for $s$ upon further increasing $\delta_\text{max}$. In order to estimate the value of $s$ for $\delta
_\text{max}\to\infty$ we perform a linear fit of $s$ against $1/\delta_\text{max}$, using $\delta_\text{max}\in\{600, 650, 700, 750, 800\}$ for $\sigma\in[0,3)$ and $\delta_\text{max}\in \{300, 350, 400, 450, 500\}$ for $\sigma\in[3,6]$ (note that for $\sigma\gtrsim 1$ this does not change $s$ significantly).

\begin{figure}[t!]
	\centering
	\includegraphics[width=1.\columnwidth]{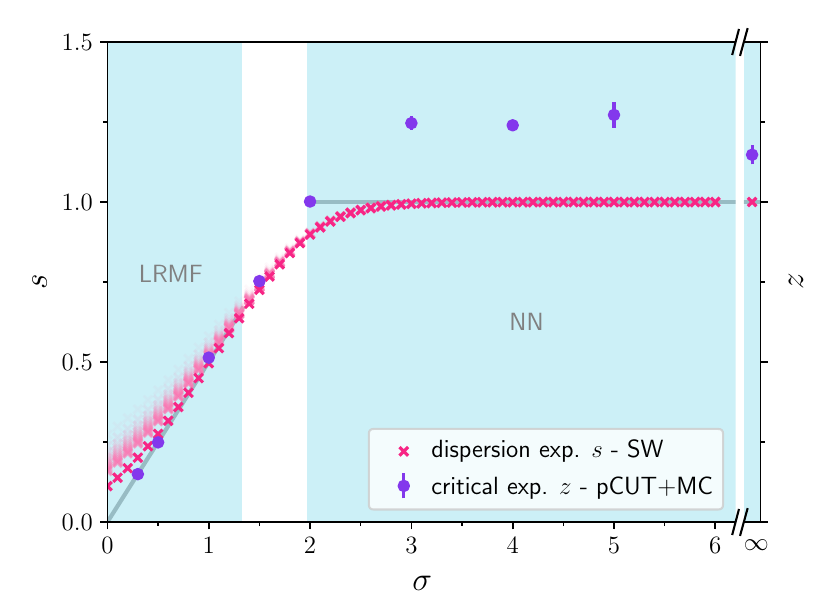}
	\caption{Exponent $s$ of the low-energy behavior of the dispersion in the antiferromagnetically order phase from spin-wave (SW) calculations and dynamical critical exponent $z$ from the pCUT+MC approach as a function of the long-range decay exponent $\sigma$ for the Heisenberg bilayer model ($\theta=\pi/4$). The NN and LRMF regimes are visualized by the blue background and the expected values of $z$ by gray lines. While the critical exponent $z$ is well above the expected values $z=1$ in the NN regime for $\sigma>2$, the exponent shows a linear behavior for $\sigma\le 2$ as expected. The exponent $s$ is exactly one in the NN regime and shows linear behavior below the NN regime except for some rounding at the boundary and deviations for very small $\sigma$ which is likely an artifact due to the slow convergence. The convergence behavior for different decay exponents with increasing cutoffs $\delta_{\text{max}}$ is depicted by data points with increasing opacity.}
	\label{fig:fig11}
\end{figure}

We plot the results in Fig.~\ref{fig:fig11}.
The results for the exponent $s$ for the dispersion are in line with previous results \cite{Diessel2023,Song2023b} for the square lattice Heisenberg model. 
In the regime $\sigma\ge 2$ we find $s=1$ and for $\sigma<2$ we find sublinear behavior $s<1$ of the Goldstone modes referred to as `anomalous' Goldstone regime. 
This behavior confirms the visual observations regarding the dispersion around the gapless point from Fig.~\ref{fig:fig10} in App.~\ref{app:swa}.
We see rounding at the boundary $\sigma=2$ and deviations from the linear behavior for small $\sigma\le0.5$ similar to other SW calculations \cite{Diessel2023, Song2023b}.
The low-energy behavior of the dispersion in the ordered phase characterized by the exponent $s$ shows striking agreement with the behavior of the critical exponent $z$ (see Fig.~\ref{fig:fig11}). 
The expected critical behavior of $z$ from the long-range $O(3)$ quantum rotor model agrees quantitatively well with $s$ except for the aforementioned rounding around $\sigma=2$ and deviations for $\sigma\le0.5$ which we attribute to the convergence with $\delta_{\text{max}}$. The results from extracting $z$  with the pCUT+MC approach agree with the behavior of $s$ quantitatively for $\sigma\le 2$ and qualitatively for $\sigma >2$.
These findings confirm that we can indeed infer the behavior of gapless Goldstone modes in the symmetry-broken phase by determining the critical exponent $z$ from extrapolating high-order series of the dispersion in the disordered phase to the critical point. 
This is possible since the low-energy properties of the dispersion at the critical point must be adiabatically connected to properties of a gapless dispersion deep in the ordered phase.

\bibliography{bibliography.bib}

\end{document}